\documentclass[a4paper,fleqn,usenatbib]{mnras}

\usepackage{newtxtext,newtxmath}
\usepackage[T1]{fontenc}
\usepackage{ae,aecompl}

\usepackage{graphicx}	% Including figure files
\usepackage{amsmath}	% Advanced maths commands
\usepackage{amssymb}	% Extra maths symbols

\usepackage{multirow, makecell}
\usepackage{booktabs,caption}
\usepackage{threeparttable}
\usepackage{hyperref}

\def\arcsec{$^{\prime\prime}$}
\def\arcmin{$^{\prime}$}

\newcommand{\NtwoH}{N$_{2}$H$^{+}$}
\def\tweCO{$^{12}$CO}

\newcommand{\HthrCO}{H$^{13}$CO$^{+}$}

\def\deg{$^{\circ}$}

\newcommand{\kms}{km~s$^{-1}$}

\title[Velocity Structures in Dense Cores with GBT-Argus]{Investigating the Complex Velocity Structures within Dense Molecular Cloud Cores with GBT-Argus}

\author[C.-Y. Chen et al.]{
Che-Yu Chen,$^{1}$\thanks{E-mail: cc6pg@virginia.edu}
Shaye Storm,$^{2}$
Zhi-Yun Li,$^{1}$
Lee G. Mundy,$^{3}$
David Frayer,$^{4}$
Jialu Li,$^{3}$
\newauthor Sarah Church,$^{5}$
Rachel Friesen,$^{6,7}$
Andrew I. Harris,$^{3}$
Leslie W. Looney,$^{8}$
Stella Offner,$^{9}$
\newauthor Eve C. Ostriker,$^{10}$
Jaime E. Pineda,$^{11}$
John Tobin,$^{6}$
and Hope H.-H.~Chen$^{9}$
\\
$^{1}$Department of Astronomy, University of Virginia,
  Charlottesville, VA 22904, USA\\
$^{2}$Harvard-Smithsonian Center for Astrophysics, Cambridge, MA 02138, USA\\
$^{3}$Department of Astronomy, University of Maryland,
  College Park 20742, MD, USA\\
$^{4}$Green Bank Observatory, Green Bank, WV 24944, USA\\
$^{5}$Physics Department, Stanford University, Stanford, CA 94305, USA\\
$^{6}$National Radio Astronomy Observatory, Charlottesville, VA 22904, USA\\
$^{7}$Department of Astronomy \& Astrophysics, University of Toronto, Toronto, ON Canada M5S 3H4\\
$^{8}$Department of Astronomy, University of Illinois, Urbana, IL 61801, USA\\
$^{9}$Department of Astronomy, University of Texas, Austin, TX 78712, USA\\
$^{10}$Department of Astrophysical Sciences, Princeton University, Princeton, NJ 08544, USA\\
$^{11}$Max Planck Institute for Extraterrestrial Physics, 85748 Garching, Germany\\
}

\date{Accepted XXX. Received YYY; in original form ZZZ}

\pubyear{2019}

\begin{document}
\label{firstpage}
\pagerange{\pageref{firstpage}--\pageref{lastpage}}
\maketitle

\begin{abstract}
%This is a simple template for authors to write new MNRAS papers. The abstract should briefly describe the aims, methods, and main results of the paper. It should be a single paragraph not more than 250 words (200 words for Letters). No references should appear in the abstract.
We present the first results of high-spectral resolution ($0.023$~\kms) \NtwoH{} observations of dense gas dynamics at core scales ($\sim 0.01$~pc) using the recently commissioned {\it Argus} instrument on the Green Bank Telescope (GBT). 
While the fitted linear velocity gradients across the cores measured in our targets nicely agree with the well-known power-law correlation between the specific angular momentum and core size, it is unclear if the observed gradients represent core-scale rotation. 
In addition, our {\it Argus} data reveal detailed and intriguing gas structures in position-velocity (PV) space for all 5 targets studied in this project, which could suggest that the velocity gradients previously observed in many dense cores actually originate from large-scale turbulence or convergent flow compression instead of rigid-body rotation.
We also note that there are targets in this study with their star-forming disks nearly perpendicular to the local velocity gradients, which, assuming the velocity gradient represents the direction of rotation, is opposite to what is described by the classical theory of star formation. This provides important insight on the transport of angular momentum within star-forming cores, which is a critical topic on studying protostellar disk formation.
\end{abstract}

\begin{keywords}
ISM: kinematics and dynamics -- ISM: molecules -- radio lines: ISM -- stars: formation -- stars: protostars
\end{keywords}

%%%%%%%%%%%%%%%%%%%%%%%%%%%%%%%%%%%%%%%%%%%%%%%%%%

%%%%%%%%%%%%%%%%% BODY OF PAPER %%%%%%%%%%%%%%%%%%

\section{Introduction}
\label{sec::intro}

Stars form in dense cores of molecular clouds (hereafter MCs). 
The classical theory describes star formation as a four-stage process: 1) the formation of dense cores within MCs by losing magnetic and turbulent support, 2) the collapse of gravitationally unstable cores and the formation of protostellar disks through angular momentum conservation, 3) the onset of stellar winds and bipolar outflows, and 4) the termination of infall and the birth of young stellar objects (YSOs) with circumstellar disks \citep{1987ARA&A..25...23S}.
%Gravity, turbulence, and magnetic field are considered the three crucial factors during this process. 
Magnetic effects, turbulence, and self-gravity are considered the three key agents affecting the dynamics of star forming process in MCs at all physical scales and throughout different evolutionary stages \citep{2007ARA&A..45..565M}.
In order to understand the properties of the stars and disks, the end product of the star formation process, one must determine the properties of the dense cores, which provide the initial conditions for star formation. 
%Dense cores therefore provide the initial conditions for star formation.
Velocity information within dense cores is therefore a key property that must be thoroughly investigated.
%because it provides the initial conditions for star formation.

Within dense cores, infall and rotation are both important in shaping the evolution of protostellar systems. 
The core's rotation rate is a critical quantity in shaping the outcome of core collapse: whether a single star or multiple system is formed \citep{2002ARA&A..40..349T}, and whether a large or small disk is produced \citep{Li_PPVI}.
The angular momentum of star-forming cores is therefore an important parameter in theoretical studies of protostellar evolution
%that has gradually attracted more attentions in theoretical studies 
\citep[e.g.][]{2010ApJ...723..425D,2018A&A...611A..24H,2018ApJ...865...34C,2019ApJ...876...33K}.
%One quantity that is especially important to determine is the core rotation rate. 
%It is known that some dense cores show a clear gradient in line-of-sight velocity, while others have a relatively random velocity field.\cite{1993ApJ...406..528G,2002ApJ...572..238C} 
In observational studies,
linearly-fitted gradient of line-of-sight velocity observed across cores is commonly used to estimate a core's angular momentum, regardless of the complex nature of the core-scale velocity field \citep{1993ApJ...406..528G}. 
Previous observations \citep[e.g.][]{1993ApJ...406..528G,2002ApJ...572..238C,2003A&A...405..639P,2007ApJ...669.1058C,2011ApJ...740...45T} as well as fully-3D MHD simulations (\citealt{2004ApJ...605..800L,2010ApJ...723..425D}; \citealt{2015ApJ...810..126C,2018ApJ...865...34C}; hereafter \hyperlink{CO15}{CO15}, \hyperlink{CO18}{CO18}) have found a power-law relationship between the specific angular momentum $J \equiv L/M$ (where $L$ is the angular momentum within the core and $M$ is the core mass) and radius $R$ for dense cores/clumps with radii $\sim~0.005-10$~pc, that $J \propto R^\alpha$ with $\alpha \approx 1.5$ (see e.g.,~\hyperlink{CO18}{CO18}).
The $J-R$ correlation over a huge range of spatial scales suggests that gas motion in cores originates at scales much larger than the core size, or the observed rotation-like features may arise from sampling of turbulence at a range of scales \citep{2000ApJ...543..822B}.
In fact, \cite{2019arXiv190605578P} recently resolved the specific angular momentum profile for three young objects in Perseus, showing an internal profile $J(r) \propto r^{1.8}$, which is between the expected values for solid-body rotation and pure turbulence.

However, dense cores often only have $\sim 0.1-0.2$~\kms{} difference in projected velocity across $\sim 0.05$~pc (see e.g.,~\hyperlink{CO18}{CO18}).
%To observationally assess the true nature of velocity structure in prestellar cores and protostellar envelopes, high velocity resolution is crucial to confirm signatures of rotation or infall, which can be limited by spectral resolution. 
High velocity and angular resolution are crucial to observationally detect a signature of rotation or infall and assess the true nature of velocity structure in prestellar cores and protostellar envelopes.
Also, the spatial dynamic range needs to be wide to cover gas motion at core/envelope scale, while also be able to spatially resolve the potentially complex structure within the core/envelope. 
%A unique combination of angular and spectral resolution, as well as the capability to map a large range of spatial scales within a reasonable amount of time, is therefore necessary for observational studies to truly characterize the kinematic features within dense cores.
A unique combination of angular and spectral resolution and fast mapping capability with full spatial scale recovery is therefore necessary for an observational study to characterize the kinematics features within dense cores.

Here we report high resolution observations using the recently-commissioned {\it Argus} focal plane array on the Green Bank Telescope (GBT), which provides an unprecedented view of core-scale velocity structures that can be directly linked to how the original core material forms and falls into the protostellar system. We observed 5 cores, both starless and protostellar, in the Perseus MC (averaged $d\approx 300$~pc; \citealt{2018ApJ...865...73O, 2019arXiv190201425Z}) with the \NtwoH{} J=1-0 line, which is a reliable tracer of cold, dense gas \citep{1997ApJ...486..316B,2002ApJ...572..238C}. We also observed \HthrCO{} in one protostellar and one starless core. \HthrCO{} is another dense gas tracer for regions where CO is not frozen out (outer envelope) or is released back into gas phase by heating from the central protostar \citep[inner envelope;][]{2011A&A...527A..88V}. We therefore use \HthrCO{} as a consistency check and focus our analysis here on \NtwoH{} observations only, unless \HthrCO{} data revealed features that are not seen in \NtwoH.

The outline of this paper is as follows. We introduce our observations in Section~\ref{sec::obs} and briefly describe our data reduction, imaging, and spectral line fitting routines in Section~\ref{sec::fitting}. The cleaned data is further investigated in Section~\ref{sec::result}, where we derive core-scale angular momentum from average velocity gradient and compare it with previous studies (Section~\ref{sec::vgrad}). Detailed velocity structures in position-velocity space are also shown (Section~\ref{sec::PV}). Further discussions and interpretations of the data for individual targets are presented in Section~\ref{sec::disc}. We summarize our results in Section~\ref{sec::sum}.

\begin{figure}
\begin{center}
\includegraphics[width=\columnwidth]{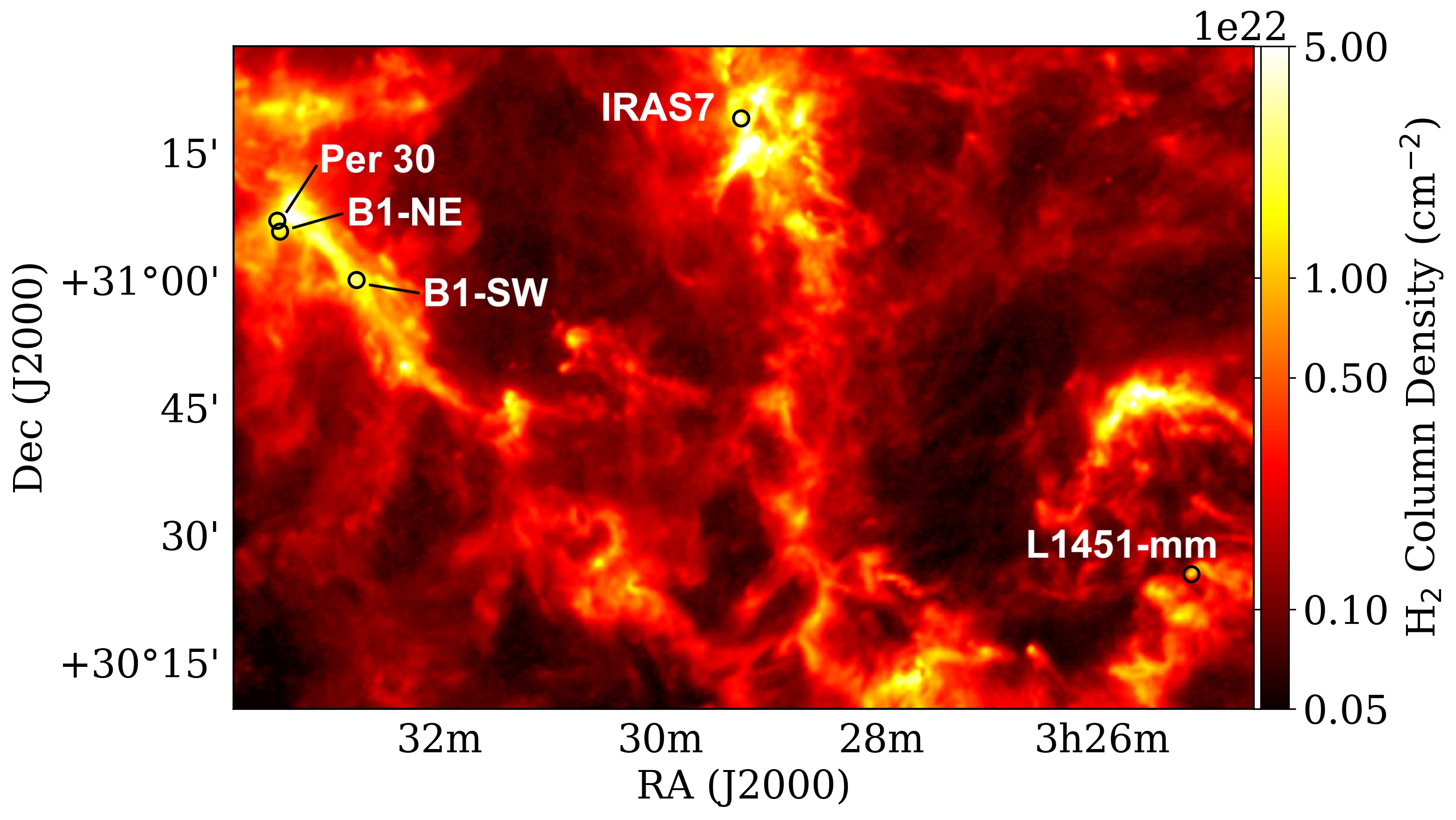}
\vspace{-.15in}
\caption{The Perseus Molecular Cloud in {\it Herschel} column density (\citealt{2010A&A...518L.102A}; Pezzuto et al. in prep.) with the locations of the 5 targets discussed in this manuscript.}
\label{Herschelmap}
\end{center}
\end{figure}

\begin{table*}
	\centering
  \begin{threeparttable}
	\caption{Summary of targets}
	\label{tab::obj}
	\begin{tabular}{l | cccccc } 
		\hline
%		\multirow{3}{*}{Core Name} & \multicolumn{6}{c|}{inner} & \multicolumn{6}{c}{inner $+$ outer} \\
%		\cline{2-13}
		 \multirow{2}{*}{target name} & \multirow{2}{*}{region} & R.A. & Dec. & $1.1$~mm  & system velocity & \multirow{2}{*}{type}  \\
		 & & (J2000) & (J2000) & source ID$^\star$ & [\kms] & \\
		\hline
		L1451-mm & L1451 & 03:25:10.247 & +30:23:49.25 & Bolo~2\ \ \  & 4.0 & Class 0  \\
		Per~30 & Barnard 1& 03:33:26.948 & +31:06:54.89 & Bolo~84 & 7.0  & Class 0/I \\
		B1-NE & Barnard 1& 03:33:24.918 & +31:05:45.92 & Bolo~82 & 6.8 & starless  \\
		B1-SW & Barnard 1& 03:32:44.367 & +31:00:03.08 & Bolo~70 & 6.8 & starless  \\
		IRAS7 & NGC~1333 & 03:29:11.949 & +31:18:30.21 & Bolo~49 & 8.5 & triple system$^\dagger$  \\
		\hline
	\end{tabular}
    \begin{tablenotes}
      \footnotesize
%      \item $^\S$Note that this is the integration time per {\it Argus} beam, and the 16 beams of {\it Argus} overlapped during the mapping. As a result, the center part of the map was observed with longer total integration time than the edge of the map.
%      \item $^*$Per~30 and B1-NE were observed together in both \NtwoH{} ($3.54$~hr) and \HthrCO{} ($3.96$~hr).
      \item $^\star$As defined in \cite{2006ApJ...638..293E}.
      \item $^\dagger$IRAS7 contains two Class 0 sources (Per~18 and Per~21) and a Class I source (Per~49).
    \end{tablenotes}
  \end{threeparttable}
\end{table*}

\section{Data}
\label{sec::data}

{\it Argus} is a 16-pixel focal plane array operating in the $85-116$~GHz range on GBT \citep{2014SPIE.9153E..0PS}, which is designed for efficient and sensitive large-area mapping. 
%For a spectral line observation over a region much larger than the telescope beam size, the mapping speed scales as $\sim n_\mathrm{pix}/(\Delta T_\mathrm{min})^2$, where $n_\mathrm{pix}$ is the number of receiving pixels and $\Delta T_\mathrm{min}$ is the line detection sensitivity. 
%As GBT provides a beam size of $\sim 8$\arcsec{} at 90~GHz, the $4\times 4$ {\it Argus} array with each receiving pixel separated by 30.4\arcsec{} on the sky is able to significantly improve the mapping speed. 
The GBT provides a beam size of $\sim 8$\arcsec{} at 90~GHz, and {\it Argus} is configured in a $4\times 4$ array with each pixel separated by 30.4\arcsec{} on the sky.
Also, {\it Argus} has a low system temperature by using advanced Monolithic Millimeter-wave Integrated Circuit (MMIC) technology \citep{MMIC_2009}, which gives receiver noise temperatures of less than 53 K per pixel.

\subsection{Targets and Observations}
\label{sec::obs} 

Five cores in Perseus were selected for this study. 
Figure~\ref{Herschelmap} shows an annotated {\it Herschel} column density map (\citealt{2010A&A...518L.102A}; Pezzuto et al. in prep.) illustrating the locations of our targets.
These cores have effective radii $\sim 20$\arcsec, or about $0.03$~pc at $d\approx 300$~pc, which is around the critcal scale in studying the evolution of angular momentum within star-forming cores where gravitational collapse normally happens \citep[for a review, see e.g.,][]{Li_PPVI}.
The targets were chosen to cover different evolutionary stages (a mix of starless and protostellar cores) and ambient environments (cluster neighborhood versus relatively isolated cores). Table~\ref{tab::obj} lists the basic properties of these 5 targets.

Observations presented here were conducted using GBT between Nov 18 2017 and Jan 30 2018 with the on-the-fly (OTF) method (see \citealt{2007A&A...474..679M} and references therein).
We used the {\it Argus} receiver to map \NtwoH{} J=1-0 and \HthrCO{} J=1-0 emission with the VEGAS backend. We started each session with observations of 3C84 to adjust the telescope surface for thermal corrections, to determine pointing corrections, and to focus for the receiver. 
We configured the VEGAS backend to a rest frequency of 93173.704~MHz or 86754.28840~MHz for \NtwoH{} or \HthrCO{} observations, respectively, using mode~6 with 187.5 MHz of bandwidth and 1.43 kHz spectral resolution. 
System temperature calibration for all 16 receiver pixels was done before observing the science targets, and we performed calibration, pointing, and focus scans every 30-50 minutes depending on the weather.
We mapped the science targets in RA and DEC scan directions. 
The data were taken every 2~seconds with map scan rates of about 0.92\arcsec{} per second; this led to angular sampling of about 1.8\arcsec{} per sample, which was about $4-5$ times less than the expected angular size of the beam. 
Frequency switching was used with offsets of $-12.5$ and $+12.5$~MHz. 
%The on-source integration time was determined using previous \NtwoH{} data in Perseus from the CARMA Large Area Star Formation Survey \citep[CLASSy;][]{2014ApJ...794..165S} to reach our requested sensitivity $\sim 0.05$~K.
The beam sizes in the calibrated images are 9.4\arcsec{} and 9.9\arcsec{} for \NtwoH{} and \HthrCO, respectively. 
%The slewing speed of the telescope: interval between each data bump; the data were taken every xx seconds. The distance between each column is xx (state that the sampling satisfying the Nyquist frequency) The noise level of each spectrum reached ~ xx.  
%2. Details of observations for each source. I would suggest adding a column in Table 1 indicating the mapping size of each source + V_lsr (it looks like some regions were mapped more than once from my quick check of Sheya's observing scripts, but probably need a more carefully check of the observing log.)
%The observations are summarized in Table~\ref{tab::obs}.

\begin{table*}
	\centering
  \begin{threeparttable}
	\caption{Summary of observations}
	\label{tab::obs}
	\begin{tabular}{l | cccccc } 
		\hline
%		\multirow{3}{*}{Core Name} & \multicolumn{6}{c|}{inner} & \multicolumn{6}{c}{inner $+$ outer} \\
%		\cline{2-13}
		 \multirow{2}{*}{target name} & on-source & \NtwoH{} $I_\mathrm{peak}$ & \NtwoH{} $I_\mathrm{bg}$ & \NtwoH{} & \HthrCO{} $I_\mathrm{peak}$ & \HthrCO{} \\
		 & time [hr]$^\star$ & [K~\kms] & [K~\kms] & sensitivity [K] & [K~\kms] & sensitivity [K] \\
		\hline
		L1451-mm & 1.64 & \ \ 4.05 & 0.43 & 0.26 & $-$ & $-$\\
		Per~30 & \multirow{2}{*}{$3.54/3.96^\dagger$} & \ \ 5.33 & 0.64 & \multirow{2}{*}{0.18} & 0.75 & \multirow{2}{*}{0.14}\\
		B1-NE & & \ \ 3.45 & 0.45 & & 0.24 & \\
		B1-SW & 2.25 & \ \ 4.36 & 0.49 & 0.25 & $-$ & $-$\\
		IRAS7 & 1.64 & 17.35 & 1.63 & 0.26 & $-$ & $-$\\
		\hline
	\end{tabular}
    \begin{tablenotes}
      \footnotesize
      \item $^\star$Note that this is the integration time per {\it Argus} beam, and the 16 beams of {\it Argus} overlapped during the mapping. As a result, the center part of the map was observed with longer total integration time than the edge of the map.
      \item $^\dagger$Per~30 and B1-NE were observed together in both \NtwoH{} ($3.54$~hr) and \HthrCO{} ($3.96$~hr).
    \end{tablenotes}
  \end{threeparttable}
\end{table*}

\subsection{Data Reduction and Spectral Fitting}
\label{sec::fitting}

{\it Argus} uses the chopper wheel method for calibration, which is the standard procedure in mm and sub-mm spectral line observation \citep{1981ApJ...250..341K}. The data is thus calibrated in ${T_a}^*$ scale, which corrects for atmospheric attenuation, resistive losses and rearward spillover and scattering. The GBT weather database returns the atmospheric temperature and opacity information. 
The efficiency of {\it Argus} is adopted to be $\eta = 50.5$\% and $50$\% for \NtwoH{} (93.17~GHz) and \HthrCO{} (86.75~GHz), respectively.
Note that this is the telescope efficiency associated with spatial scale of Jupiter, which is slightly larger than the primary main-beam efficiency of small, point-like sources (see the GBT Memo \#302: \citealt{2019arXiv190602307F}). The overall uncertainty on the flux is $\sim 15$~\%. Nevertheless, we caution the readers that full characterization is still ongoing, and the absolute amplitude uncertainty is not used for any of the discussion below, only statistical uncertainties.
%$\eta_{MB}$ = 0.5 has been temporarily taken here to convert the data to T$_{MB}$ scale. 
%Data reduction of this project was processed via GBTIDL, including the calibration and gridding 

We performed standard calibration using \texttt{GBTIDL} including baseline subtraction, and
%We then ran \texttt{gbtgridder} to generate position-position-frequency data cubes with pixel size $\sim 2$~\arcsec, 
\texttt{gbtgridder} was used to make data cubes with pixel size 2\arcsec~$\times$~2\arcsec{} from maps per frequency using a Gaussian kernel. 
%The data cubes were later converted to position-position-velocity (PPV) cubes using Python package \texttt{SpectralCube}.
%with \NtwoH{} $J = 1\rightarrow 0$ main line rest frequency $93.1737637$~GHz \citep{2009A&A...494..719P}, and \HthrCO{} rest frequency $86.754294$~GHz.
When re-gridding the data, we chose to combine 5 and 10 channels for \NtwoH{} and \HthrCO{} to reach velocity resolution $\approx 0.023$ and $0.05$~\kms, respectively. The sensitivity of our final \NtwoH{} maps is $\sim 0.15-0.25$~K, and $\sim 0.14$~K for the \HthrCO{} observation (see Table~\ref{tab::obs}).
We used the Python package \texttt{PySpecKit} \citep{2011ascl.soft09001G} for spectral fitting, which simultaneously fits the 15 hyperfine lines of \NtwoH{} J=1-0 and returns the fitted centroid velocity, linewidth, excitation temperature, and optical depth. We adopted a signal-to-noise cut to peak line intensity of $S/N > 5$ for the \NtwoH{} data ($S/N > 3$ for \HthrCO) when performing the fitting.

\begin{figure*}
\begin{center}
\includegraphics[width=\textwidth]{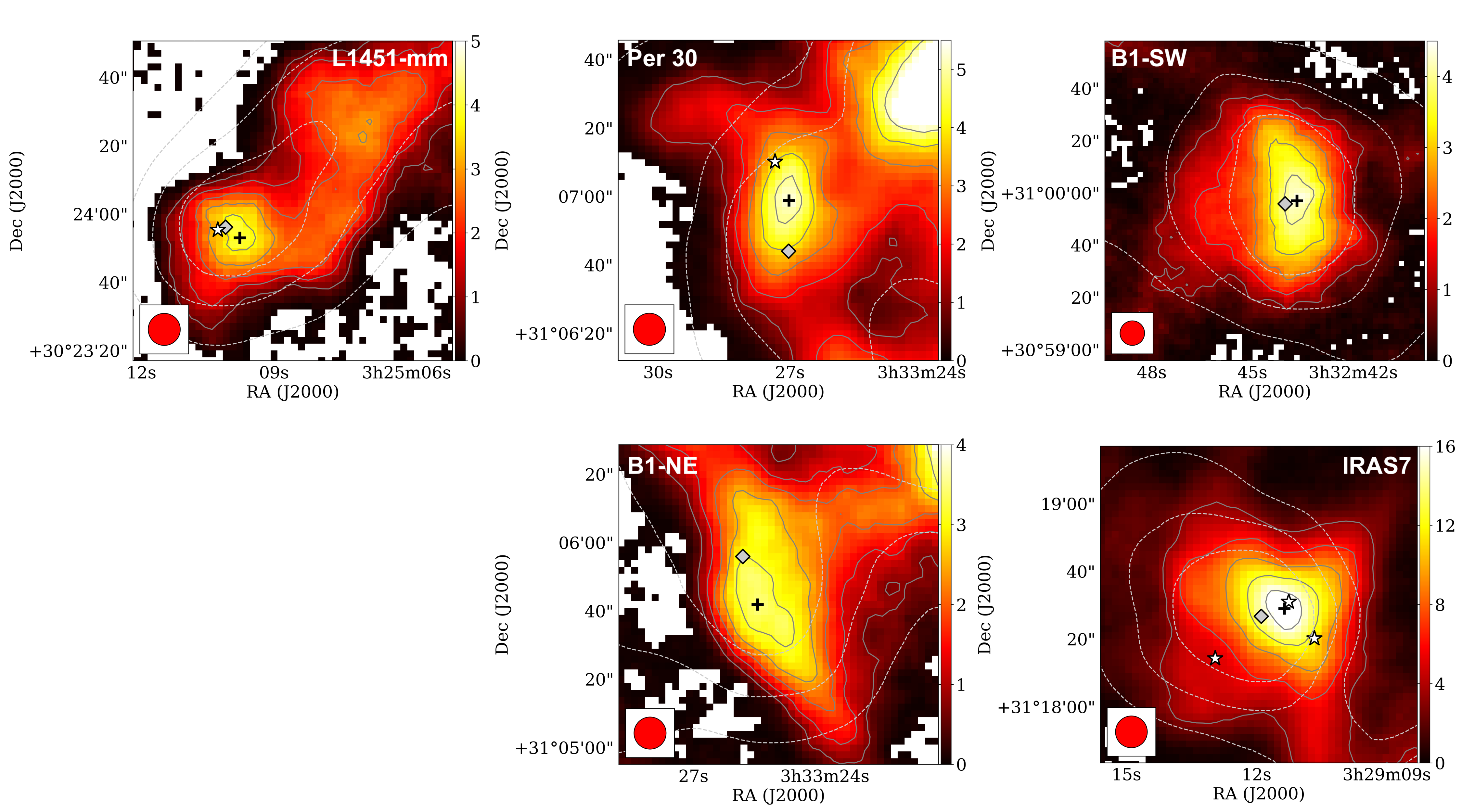}
\vspace{-.15in}
\caption{{\it Argus} \NtwoH{} integrated intensity maps in K~\kms{} (with $S/N > 2$ masks on PPV data cubes) of FHSC-candidate L1451-mm ({\it top left}), protostellar core Per~30 ({\it top middle}), prestellar cores B1-SW ({\it top right}) and B1-NE ({\it bottom middle}), and triple system IRAS7 ({\it bottom right}). 
Note that the colorscale of IRAS7 differs significantly from other panels.
Known protostars ({\it white stars}; \citealt{2009ApJ...692..973E,2016ApJ...818...73T}) and peak locations of {\it Herschel} column density ({\it grey diamonds}) and \NtwoH{} emission ({\it black cross}) are also marked. 
The contour levels are $10\%$, $30\%$, $50\%$, $70\%$, $90\%$ of $(I_\mathrm{peak} - I_\mathrm{bg})$ above $I_\mathrm{bg}$ for \NtwoH{} integrated intensity ({\it grey contours}), and $25\%$, $50\%$, $75\%$ of $(I_\mathrm{peak} - I_\mathrm{bg})$ above $I_\mathrm{bg}$ for {\it Herschel} column density ({\it dashed contours}).
}
\label{mom0}
\end{center}
\end{figure*}

\section{Results}
\label{sec::result}

Figure~\ref{mom0} summarizes our observations by showing the \NtwoH{} integrated intensity maps of the observed cores.
For each core, we estimate the background integrated brightness $I_\mathrm{bg}$ by sampling several intensity profiles (either along RA or Dec) near the peak intensity $I_\mathrm{peak}$ and fitting each of them to a Gaussian function $G(x) = A\exp{[-(x-x_0)^2/(2\sigma^2)]}$ using the \texttt{modeling} function in Astropy. The background integrated brightness for the $i^\mathrm{th}$ profile is therefore defined as $I_\mathrm{bg}(i) = G_i(x\pm2\sigma)$, and we use the median value over all $I_\mathrm{bg}(i)$ as the background integrated brightness for the core. 
A similar procedure was applied on the {\it Herschel} column density data as well. These measured peak and background values are then used to draw contours in moment 0 and moment 1 maps like Figures~\ref{mom0}$-$\ref{mom1IRAS7} (and to define core boundaries; see Sections~\ref{sec::vgrad} and \ref{sec::PV} below).
%in Figure~\ref{mom0}, we show contours at $10\%$, $30\%$, $50\%$, $70\%$, $90\%$ of $(I_\mathrm{peak} - I_\mathrm{bg})$ above $I_\mathrm{bg}$ for \NtwoH{} integrated intensity (dark grey lines), and $25\%$, $50\%$, $75\%$ of $(I_\mathrm{peak} - I_\mathrm{bg})$ above $I_\mathrm{bg}$ for {\it Herschel} column density (dashed light grey lines).

Table~\ref{tab::obs} lists the peak and background integrated brightness of \NtwoH{} for each target.
As a sanity check, we compared the measured peak intensity and system velocity with previous studies \citep[e.g.,][]{2011ApJ...743..201P,2014ApJ...794..165S,2016ApJ...830..127S} when possible, and found that our results are consistent with those data (after correcting for the different distances that may have been assumed). Also, one of our target, B1-SW, has also been identified as a dense core by \cite{2007ApJ...668.1042K} using \NtwoH{} (their core number 79), and the measured system velocity ($6.8$~\kms) and peak integrated intensity ($\sim 4$~K~\kms) in our observation agree very well with their result.

\subsection{Linear Velocity Gradient}
\label{sec::vgrad}

\begin{figure*}
\begin{center}
\includegraphics[width=\textwidth]{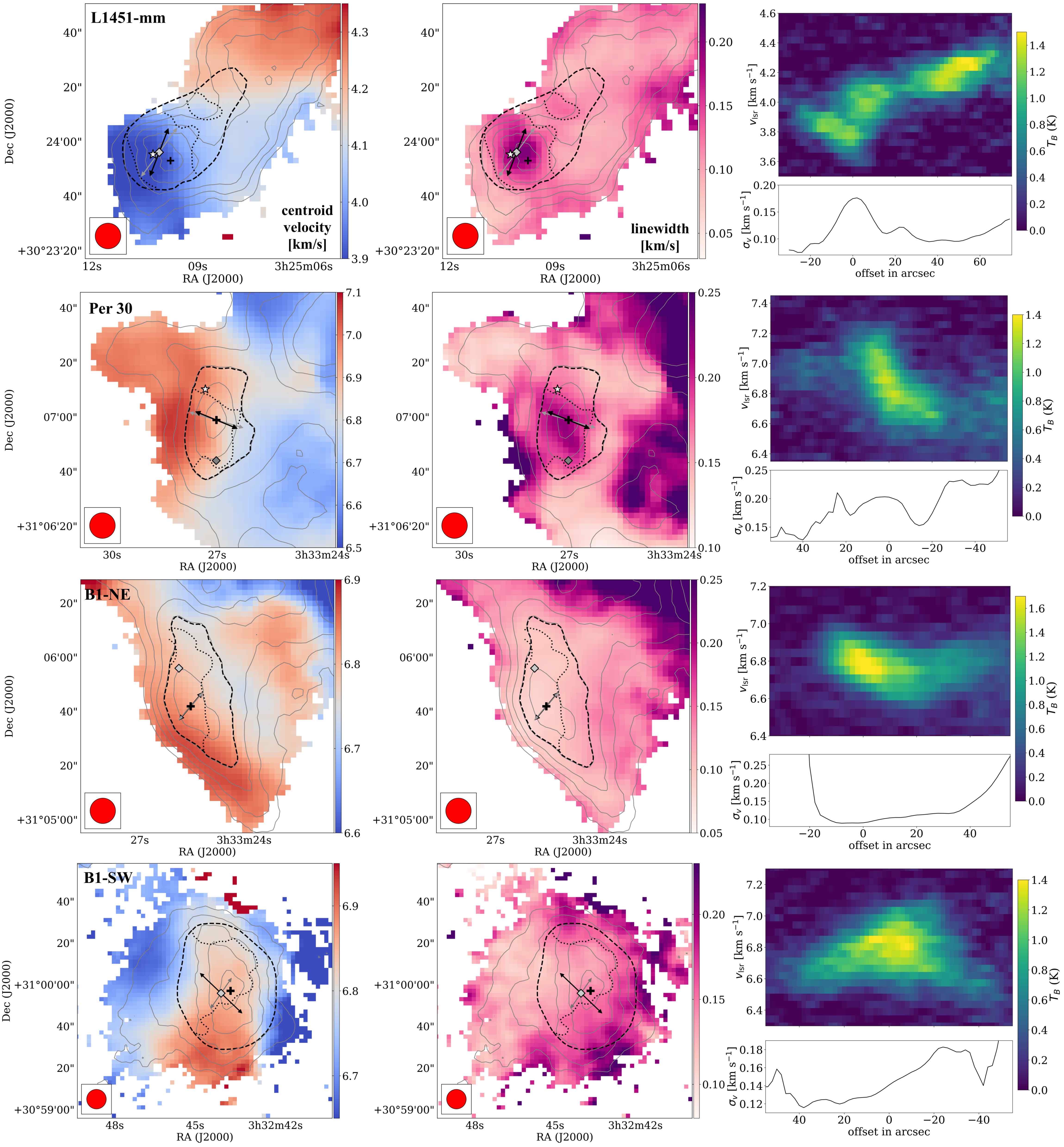}
\vspace{-.15in}
\caption{Maps of \NtwoH{} centroid velocity ({\it left column}) and linewidth ($\sigma_v$; {\it middle column}) with \NtwoH{} integrated intensity contours at the same levels as in Figure~\ref{mom0}, as well as the position-velocity (PV) diagrams (from the isolated \NtwoH{} hyperfine line only) and position-linewidth (P-$\sigma_v$) plots ({\it right column}), for our targets L1451-mm ({\it top row}), Per~30 ({\it second row}), B1-NE ({\it third row}), and B1-SW ({\it bottom row}). Arrows represent the local velocity gradients averaged over the defined core area ({\it thick black dashed contours}; see Table~\ref{tab::vgrad} for definition), both with ({\it black}) or without ({\it grey}) the $\sigma_v < \langle \sigma_v \rangle$ mask ({\it dotted contours}). The PV diagrams and P-$\sigma_v$ plots are drawn within a 20\arcsec-wide zone along the black arrows, with offsets measured from the peak locations of either the {\it Herschel} column density ({\it grey diamonds}) or the \NtwoH{} emission ({\it black cross}), depending on which one was used to define core boundaries. Known protostars are also marked as white stars.}
\label{mom1}
\end{center}
\end{figure*}

\begin{figure*}
\begin{center}
\includegraphics[width=\textwidth]{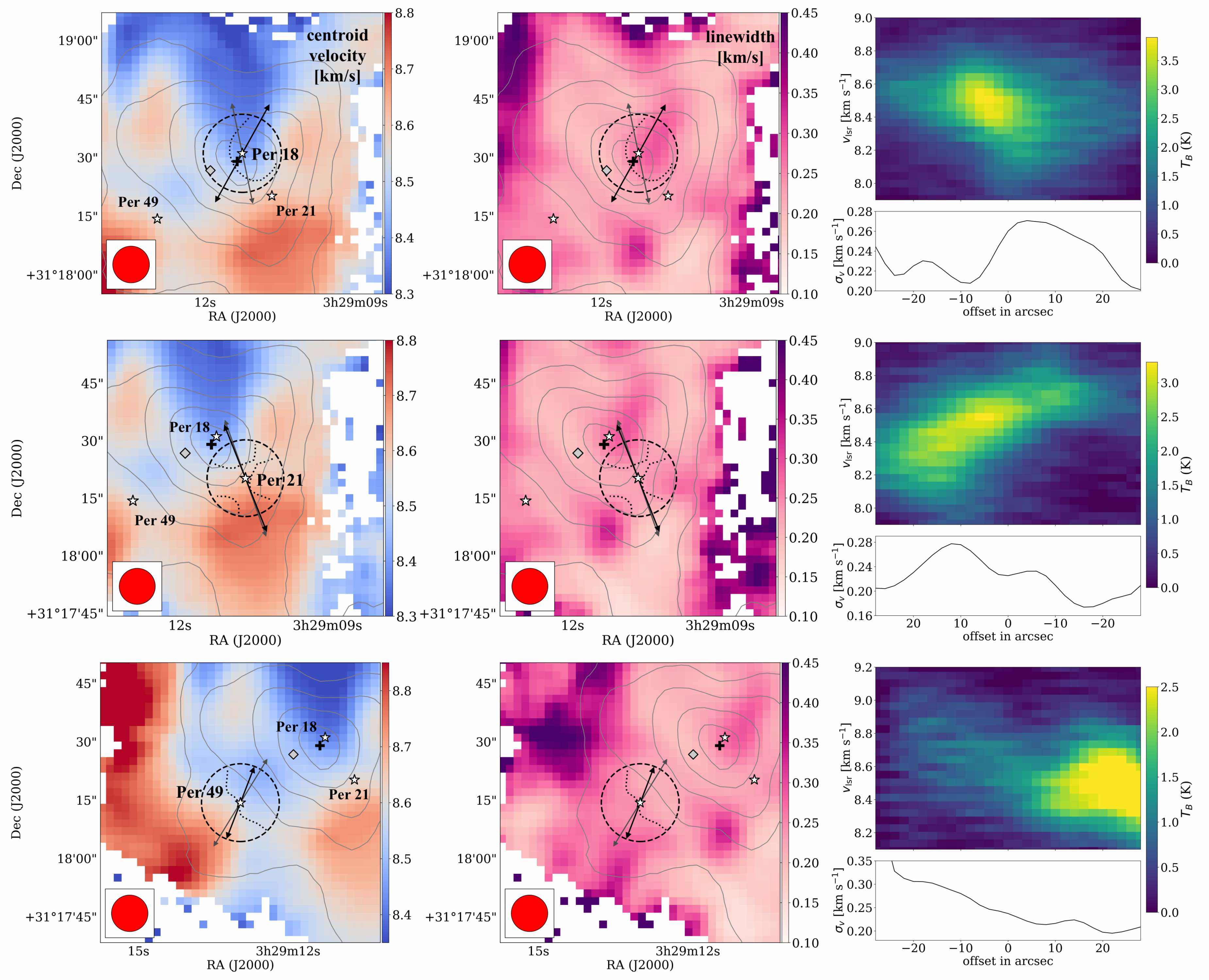}
\vspace{-.15in}
\caption{Same as Figure~\ref{mom1} but with 10\arcsec-radius circles as the core boundaries ({\it dashed circles}; see text), for individual protostellar systems in IRAS7: Per~18 ({\it top row}), Per~21 ({\it central row}), and Per~49 ({\it bottom row}). The PV diagrams and P-$\sigma_v$ plots are drawn within a 10\arcsec-wide zone along the black arrows.}
\label{mom1IRAS7}
\end{center}
\end{figure*}

\begin{table*}
	\centering
  \begin{threeparttable}
	\caption{Linear velocity gradients measured in \NtwoH.}
	\label{tab::vgrad}
	\begin{tabular}{l | crlrlccr } 
		\hline
%		\multirow{3}{*}{Core Name} & \multicolumn{6}{c|}{inner} & \multicolumn{6}{c}{inner $+$ outer} \\
%		\cline{2-13}
		 \multirow{2}{*}{Core} & core & \multicolumn{2}{c}{$\nabla v_\mathrm{lsr}$ (without $\langle\sigma_v\rangle$ mask)$^\S$} & \multicolumn{2}{c}{$\theta_{\nabla v_\mathrm{lsr}}$ (without $\langle\sigma_v\rangle$ mask)$^\dagger$} & $r_\mathrm{core}$ & \multirow{2}{*}{$(b/a)_\mathrm{core}$} & $\theta_a$$^\dagger$\\
		 & definition & \multicolumn{2}{c}{[\kms{} pc$^{-1}$]} & \multicolumn{2}{c}{[\deg]} & [pc] & &  [\deg] \\
		\hline
		\multirow{2}{*}{L1451-mm} & Herschel 75\% & $3.03\pm 1.77$ & ($3.72\pm 2.17$) & $-24.1$ & $(\ \ -40.5)$ & 0.031 & 0.51 & $-54.9$ \\
		& \NtwoH{} 70\% & $4.62\pm1.50$ & $(5.10\pm1.83)$ & $-76.3$ & $(\ \ -71.7)$ & 0.015 & 0.81 & $-76.1$  \\
		Per~30 & \NtwoH{} 50\% & $3.03\pm 1.05$ & $(3.75\pm 1.52)$ & 70.9 & $(\ \ \ \ 76.2)$ & 0.025 & 0.67 & $-7.3$ \\
		B1-NE & \NtwoH{} 70\% & $1.39\pm 0.65$ & $(1.36\pm 0.42)$ & 134.6 & $(\ \ 135.5)$ & 0.028 & 0.49 & $18.3$  \\
		B1-SW & Herschel 75\% & $1.93\pm 1.59$ & $(1.19\pm 2.41)$ & $-129.3$ & $(\ \ 141.2)$ & 0.043 & 0.86 & $27.0$ \\
%		\hline
		IRAS7 Per~18 & $10$\arcsec{} circle & $7.09\pm 2.68$ & $(6.28\pm 4.25)$ & $147.0$ & $(-166.3)$ & 0.015 & $-$ & $-$ \\
		IRAS7 Per~21 & $10$\arcsec{} circle & $7.31\pm 2.29$ & $(7.81\pm 2.70)$ & $-155.1$ & $(-157.5)$ & 0.015 & $-$ & $-$ \\
		IRAS7 Per~49 & $10$\arcsec{} circle & $4.79\pm 2.14$ & $(6.58\pm 3.30)$ & $155.7$ & $(\ \ 144.3)$ & 0.015 & $-$ & $-$  \\
		\hline
	\end{tabular}
    \begin{tablenotes}
      \footnotesize
      \item $^\S$Both $\nabla v_\mathrm{lsr}$ values with ({\it left column}) and without ({\it right column in ()}) the $\sigma_v < \langle\sigma_v\rangle$ mask (see text) are listed here in the format of ({\it mean value})~$\pm$~({\it standard deviation}).
      \item $^\dagger$Measured counterclockwise toward east from north.
    \end{tablenotes}
  \end{threeparttable}
\end{table*}

Figures~\ref{mom1} and \ref{mom1IRAS7} summarize the kinematic features observed in our targets by showing maps of the line-of-sight velocity (left panels), linewidth (middle panels), and position-velocity diagrams (right panels; will be discussed in Section~\ref{sec::PV}) of these targets.
% describe vlsr and sigma maps
Generally speaking and considering the typical gas temperature $\sim 10$~K ($c_s\approx 0.2$~\kms), almost all cores show smooth velocity structures with sub- to trans-sonic linewidths inside the cores.
These zones of subsonic turbulence are usually labeled coherent cores \citep[see e.g.,][]{1998ApJ...504..223G, 2010ApJ...712L.116P, 2015Natur.518..213P}.
The only exception is the triple system IRAS7 (see Figure~\ref{mom1IRAS7}), which has a more complex velocity field at the core scale and a larger linewidth (trans- to supersonic). This implies that the IRAS7 core is more turbulent and could be a hint of turbulence-induced fragmentation within star-forming cores \citep[see e.g.,][]{2010ApJ...725.1485O,2016ApJ...827L..11O,2015Natur.518..213P}. Further discussions on IRAS7 are included in Section~\ref{sec::iras7}.

\begin{figure}
\begin{center}
\includegraphics[width=\columnwidth]{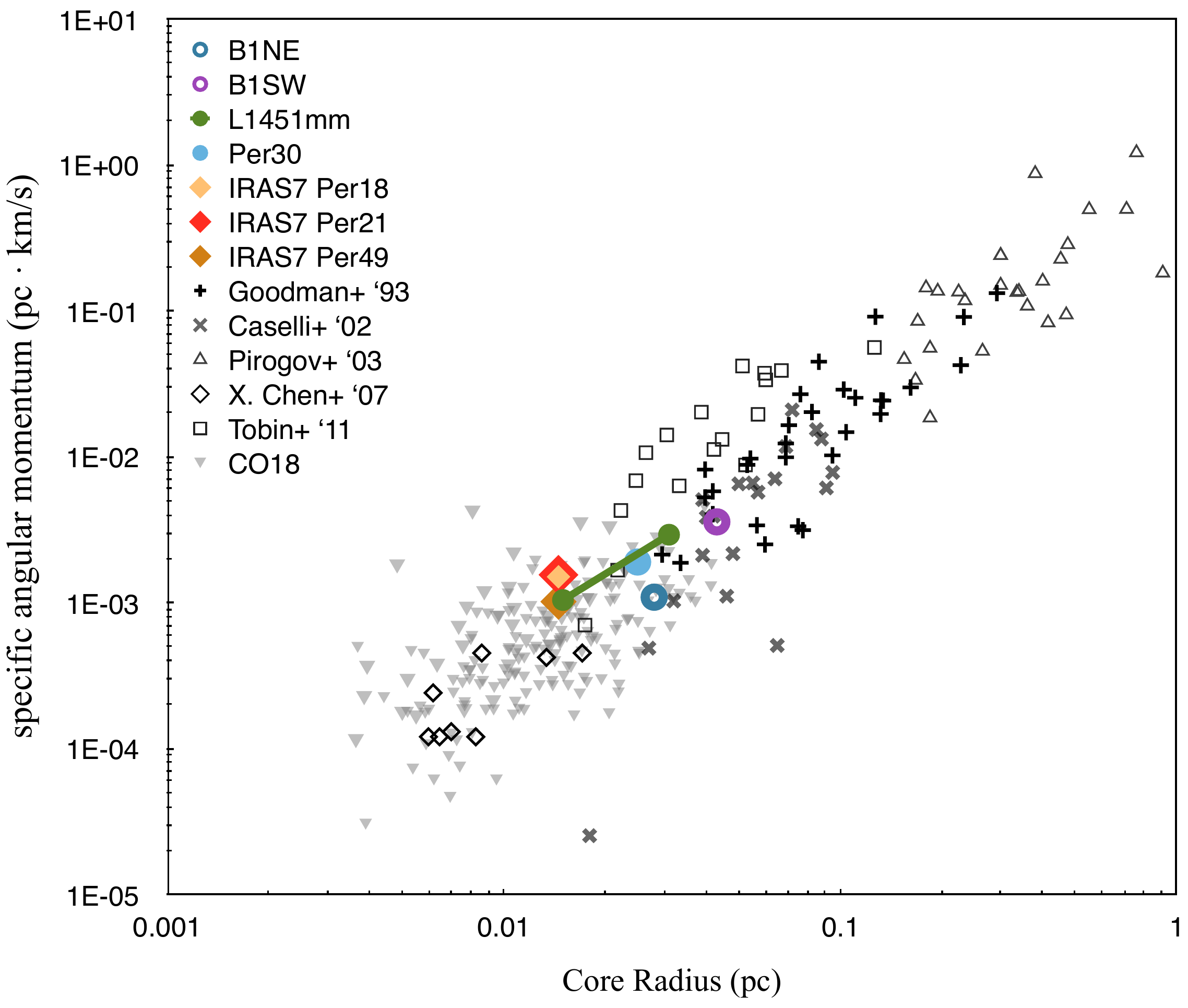}
\vspace{-.15in}
\caption{The specific angular momentum$-$radius ($J-R$) correlation measured in both previous observations and simulations ({\it black/grey symbols}), with the results of individual targets derived from this study ({\it colored symbols}), adapted from \protect\hyperlink{CO18}{CO18}. Colored open circles represent starless cores, while filled circles/diamonds represent protostellar sources. Protostars formed within the triple system IRAS7 are marked as diamonds.}
%\vspace{-.2in}
\label{JRcorr}
\end{center}
\end{figure}

As a first approach of analyzing the observed kinematic features, 
we consider the averaged linear velocity gradient at the core scale, $\nabla v_\mathrm{lsr}$.
%using only the \NtwoH{} data, 
Note that we calculated the gradient of $v_\mathrm{lsr}$ at each map pixel first and averaged over the defined core region to derive $\nabla v_\mathrm{lsr}$, which is slightly different from the 2D linear fitting procedure adopted in previous studies \citep[e.g.,][]{1993ApJ...406..528G,2002ApJ...572..238C,2011ApJ...740...45T}. 
When possible, we define core boundaries based on {\it Herschel} column density contours, since dust column density can in principle better represent the distribution of core material, not just the dense gas traced by \NtwoH. However, in cases that {\it Herschel} data failed to return a meaningful closed contour around the core area,\footnote{Targets Per~30 and B1-NE are too close to each other and thus the resolution of {\it Herschel} is not good enough to well-separate them. Similarly, the triple system IRAS7 is not resolved in the {\it Herschel} column density map.} we adopt \NtwoH{} contour levels to define cores.
We would like to point out that these are indeed very rough estimates of the core boundaries, and thus are adopted for deriving the $J-R$ correlation only (see Figure~\ref{JRcorr} and related discussions below). 

Table~\ref{tab::vgrad} lists the velocity gradient measured from our targets, as well as the contour levels adopted as core boundaries.
Each core is fitted by an ellipse to determine the effective radius of the core, $r_\mathrm{core} \equiv \sqrt{a_\mathrm{core} b_\mathrm{core}}$, where $a_\mathrm{core}$ and $b_\mathrm{core}$ are the semi-lengths of the major and minor axes of the fitted ellipse, respectively. 
We note that, though the core boundaries and thus radii are only roughly defined, there are noticeable dependences of core shape and size on the environment: cores formed within filamentary structures (e.g.,~L1451-mm) tend to be more elongated (see the minor/major axes ratio $b/a$ in Table~\ref{tab::vgrad}), while relatively isolated cores (e.g.,~B1-SW and IRAS7) are more rounded and slightly bigger in size (see e.g.,~the moment 0 maps in Figure~\ref{mom0}).

The position angle of the core is also listed in Table~\ref{tab::vgrad} as the position angle of the major axis $\theta_a$ counterclockwise from north.
The only exception is IRAS7, which is a known triple system consisting three protostars Per~18, Per~21, and Per~49. We therefore measure the velocity gradients of these three sources separately by considering circles with radius $10$\arcsec{} (a size $\sim 2\times$ of the beam) centered at these protostars as their core boundaries. 
Also, for L1451-mm we adopted two different definitions of the core, which we will discuss further in Section~\ref{sec::l1451mm}.
Nevertheless, we would like to caution the readers that though our regridded data has pixel size of 2\arcsec{}$\times 2$\arcsec, our beam size is indeed 9\arcsec. This means we might only have two independent pointings across the narrowest regions of the cores, which could impact our measurements of velocity gradient. This is a common limitation of single-dish telescopes that can be alleviated with interferometers. 

The linear velocity gradient at the core scale is defined as the average value of the gradient of $v_\mathrm{lsr}$ within the core.
Since 
%the linewidth is an indicator of whether a dominant velocity component exists, 
the fitted centroid velocity could be less accurate if the linewidth is broad,\footnote{A broad linewidth could indicate the existence of unresolved multiple velocity components along the line of sight, or potential opacity broadening if the line is optically thick.}
we adopted a mask to only include pixels with velocity dispersion smaller than the median value within the core, $\sigma_v < \langle\sigma_v\rangle_\mathrm{core}$, in determining the average linear velocity gradient for such a core. 
The median value is adopted as the selection criterion because it guarantees exactly
half of the pixels within the core boundary are used, and the statistics remains relatively unaffected among cores with or without the mask.
Though this selection criterion might remove the central part of the core from calculating $\nabla v_\mathrm{lsr}$ as in the cases of L1451-mm, Per~30, and Per~18, we note that the purpose of finding the linear velocity gradient is to derive the angular momentum of the core, while the central part of the core usually contains more infall motions (see e.g.,~Sec.~\ref{sec::l1451mm}) that could potentially contaminate the measurement of core rotation.

Both values of velocity gradient with and without such masks are listed in Table~\ref{tab::vgrad}, and their directions are illustrated in Figures~\ref{mom1} and \ref{mom1IRAS7} (black and grey arrows, respectively). 
One can easily see that in some cores, these linear velocity gradients are not good representatives of the overall velocity structure. The most obvious example is B1-SW, in which the derivation of linear velocity gradient with or without the $\sigma_v < \langle\sigma_v\rangle$ mask returns two directions that are almost perpendicular to each other (see the bottom row of Figure~\ref{mom1}). 
Note that the masking could also be considered a sanity check of how well the linear velocity gradient can be defined. If the masking introduces significant differences in the derived amplitude and direction of the velocity gradient (e.g., the case of B1-SW), it means the core-scale velocity structure is not best described by a linear, monotonic velocity field.

% J-R plot and discussion

%Regardless of its significance, 
Under the usual assumption that the velocity gradient traces rotation, 
we can use it and the fitted core size to calculate the specific angular momentum, $J_\mathrm{core} \equiv L_\mathrm{core}/M_\mathrm{core} = {r_\mathrm{core}}^2 \nabla v_\mathrm{lsr}$ \citep[see e.g.][]{1993ApJ...406..528G},\footnote{Note that we did not consider the relative angle between the velocity gradient and core shape here, as $r_\mathrm{core}$ is simply the geometric mean of the lengths of the major and minor axes; $J_\mathrm{core}$ is therefore less accurate when the aspect ratio of the core, $(b/a)_\mathrm{core}$, is much smaller than 1 (see Table~\ref{tab::vgrad}).} and compare it with the known correlation between $J$ and core size. The results are shown in Figure~\ref{JRcorr}, where we plot measurements from previous observations \citep{1993ApJ...406..528G,2002ApJ...572..238C,2003A&A...405..639P,2007ApJ...669.1058C,2011ApJ...740...45T} as well as numerical simulations (\hyperlink{CO15}{CO15}, \hyperlink{CO18}{CO18}) to compare with our data.\footnote{Note that the angular momenta of simulated cores in \hyperlink{CO18}{CO18} were measured in 3D, not projected 2D values.}
Strikingly (but not surprisingly), the specific angular momenta derived from our GBT-Argus data fit within the previous results and the known $J-R$ trend ($J\propto R^\alpha$, $\alpha\approx 1.5$) very well.
However, we would like to remind the readers that though the $J-R$ scatter plot appears to be a nice power-law correlation, it also has large uncertainties for a given radius (a factor of $\sim 3-5$) or a given value of $J$ (a factor of $\sim 10$). This also means that the uncertainties in our definition of background emission and/or core boundaries will not affect the general conclusion from our comparison.

Together with the facts that 1) rotation is not the dominant motion in the simulated cores as reported in \hyperlink{CO18}{CO18}, and 2) the internal velocity fields in our observed cores could be more complicated than having simple, monotonic gradients (see the PV diagrams on the right panels of Figures~\ref{mom1} and \ref{mom1IRAS7}; also see Section~\ref{sec::PV} below), interpreting linear velocity gradient across the core as angular momentum from rigid-body rotation may not be appropriate. In fact, as discussed in \hyperlink{CO18}{CO18}, the correlation $J \propto R^{1.5}$ can be simply treated as $\nabla v \sim \Delta v / R \propto R^{-0.5}$, or $\Delta v \propto R^{0.5}$, which agrees with the well-known property of cloud-scale turbulence in both observations and simulations \citep[as reviewed by, e.g.,~][]{2007ARA&A..45..565M}.
To summarize, although the so-called $J-R$ correlation may be misleading, since $J$ as calculated here is a proxy for angular momentum not necessarily the true angular momentum, this correlation is still real and provides a simple way to characterize a gross property of the observed velocity field. This correlation is based on directly observable quantities and should provide constraints on numerical simulations of dense cores in molecular clouds, independent of whether or not it accurately represents the true {\it specific angular momentum}$-$size correlation.

\subsection{Position-Velocity Diagrams}
\label{sec::PV}

The enhanced sensitivity of GBT with {\it Argus} made it possible to achieve an exceptionally high spectral resolution ($0.023$~\kms) in our observations, which is crucial in creating well-resolved position-velocity (PV) diagrams.
The PV diagrams of individual cores, as well as the position-averaged linewidth (P-$\sigma_v$) plots, are shown in the right panels in Figures~\ref{mom1} and \ref{mom1IRAS7}, which were drawn within a 20\arcsec-wide zone along the direction of the averaged velocity gradient (black arrows in the $v_\mathrm{lsr}$ and $\sigma_v$ maps in the same row), with offsets measured from either the peak values of {\it Herschel} column density or \NtwoH{} emission (whichever was used to define the core boundary).
We include the P-$\sigma_v$ plots here because the linewidth contains important information on whether or not the fitted centroid velocity at each sightline is well-defined. 
These PV diagrams and P-$\sigma_v$ plots clearly show that the velocity fields within these cores are significantly more complicated than simple rigid-body rotation \citep[see e.g.,][]{2012ApJ...748...16T}. In Section~\ref{sec::disc} below, we discuss in detail the observed dynamic properties of individual cores.

\section{Individual Cores}
\label{sec::disc}

\subsection{L1451-mm}
\label{sec::l1451mm}

L1451-mm (also known as Per-Bolo-2; \citealt{2006ApJ...638..293E}) is a low-mass dense core with no point-source detection in the mid-infrared \citep[$\sim 3-24~\mu$m;][]{2006ApJ...645.1246J,2007ApJS..171..447R} and very low luminosity in radio wavelengths \citep{2006ApJ...638..293E}. It has been classified as either a first hydrostatic core (FHSC) candidate \citep{2011ApJ...743..201P,2017ApJ...838...60M}, or an extremely young Class 0 protostar \citep{2016ApJ...818...73T}.

Though previous observations of L1451-mm only covered the brightest core region around the \NtwoH{} peak location, our GBT-{\it Argus} data shows that both the dense gas and velocity gradient smoothly extend to a much larger scale beyond the dense core itself (see Figure~\ref{mom1}). Also, the core boundary defined from {\it Herschel} column density contains a much larger region than the usually-considered core area based on dense gas tracers only \citep[see e.g.,][]{2011ApJ...743..201P,2017ApJ...838...60M}, indicating that the central core is not isolated from the background gas.

\begin{figure*}
\begin{center}
\includegraphics[width=0.8\textwidth]{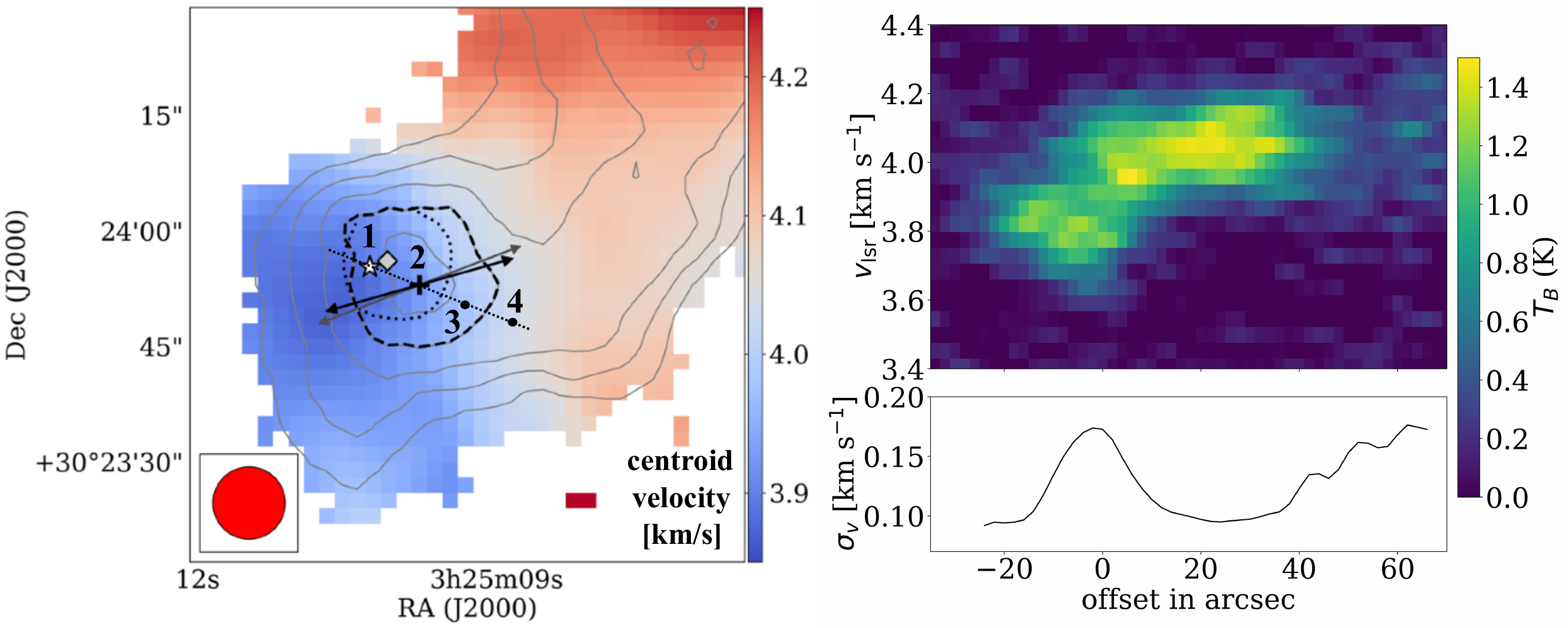}
%\vspace{-.15in}
\caption{The centroid velocity map ({\it right panel}) of L1451-mm, same as the top-left panel of Figure~\ref{mom1}, but showing velocity gradients ({\it arrows}) derived from the \NtwoH-defined core area ({\it dashed contour}) and the corresponding PV diagram and P-$\sigma_v$ plot ({\it right panel}) measured along the black arrow with respect to the \NtwoH{} peak ({\it black cross}). Numbers $1-4$ marked the equally-separated locations of spectra shown in Figure~\ref{L1451spec}.}
\label{L1451n2hp}
\end{center}
\end{figure*}

\begin{figure}
\begin{center}
\includegraphics[width=\columnwidth]{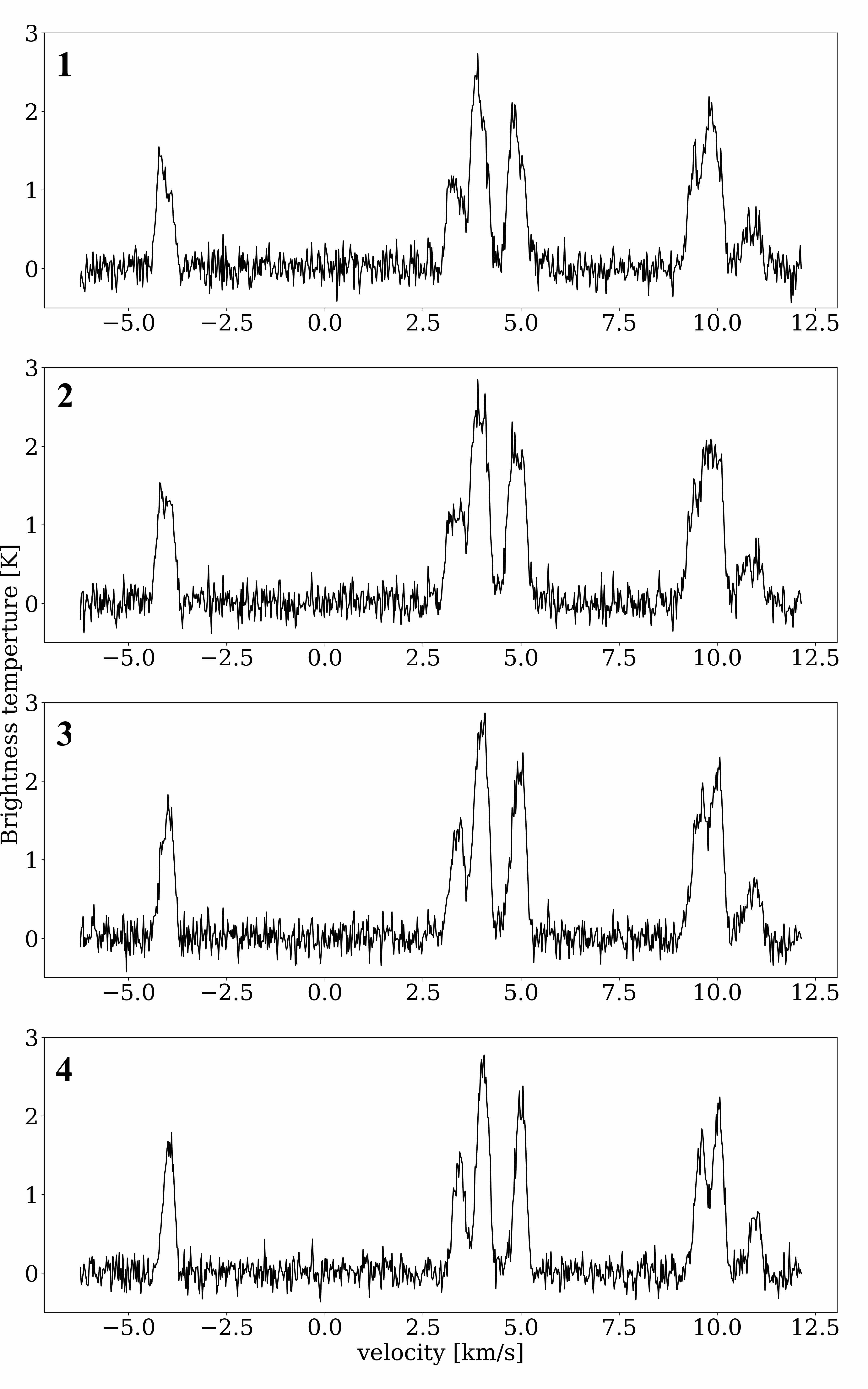}
\vspace{-.15in}
\caption{Spectra of L1451-mm at four equally-separated locations as marked in the left panel of Figure~\ref{L1451n2hp}, averaged over a beam. There are clearly two velocity components near the protostar ({\it location $1$; top panel}) and the \NtwoH{} peak ({\it location $2$; second panel}) comparing to the simple, one component line profile at locations further away from the dense core ({\it location $4$; bottom panel}).}
\label{L1451spec}
\end{center}
\end{figure}

To compare with previous studies that considered the \NtwoH{} core only, we perform the same analysis as described in Section~\ref{sec::vgrad} on L1451-mm using the \NtwoH{} $70\%$ contours (see definition in Section~\ref{sec::result}) as the core boundary. The resulting velocity gradient and PV diagram are shown in Figure~\ref{L1451n2hp},
%The amplitude and direction of the fitted linear velocity gradient are included in Table~\ref{tab::vgrad}, 
and the measured core size and derived specific angular momentum are included in the $J-R$ correlation plot, Figure~\ref{JRcorr}. Though the direction of the velocity gradient with the \NtwoH{} core is $30^\circ-50^\circ$ offset from that calculated within the {\it Herschel} core (see Table~\ref{tab::vgrad}), the two angular momenta measured at two different scales within L1451-mm follow the power-law $J-R$ correlation very well (see the two connected green dots in Figure~\ref{JRcorr}). 
This again suggests that the observed gradient in centroid velocity (which could be turbulence within the core) could be interconnected with gas motion at larger scales.

In addition, the extended emission toward the north-west part of the dense core can be clearly seen in the PV diagram in Figure~\ref{L1451n2hp}, which further indicates that the \NtwoH{} core is dynamically associated with the surrounding gas.
We note that this is similar to the observations reported by \cite{2011A&A...533A..34H}, who found that gas velocities traced by \NtwoH{} and C$^{18}$O (which in principle traces less dense gas compared to \NtwoH) agree with each other very well.
Though we do not have larger scale information and multiple density tracers as in \cite{2011A&A...533A..34H}, the fact that gas kinematics within the central part of the core is continuous from the surrounding gas in our observation also indicates the connection between the motions within cores and at larger scales.

More importantly, from both PV diagrams (in Figures~\ref{mom1} and \ref{L1451n2hp}) we note that there clearly are two velocity components near either the protostar or the \NtwoH{} peak, which could be the reason that the linewidth is significantly higher inside the \NtwoH{} core than the surrounding gas (see the $\sigma_v$ map and P-$\sigma_v$ plot in Figure~\ref{mom1}). 
Four sample spectra (at locations marked as numbers $1-4$ on Figure~\ref{L1451n2hp}) are shown in Figure~\ref{L1451spec}, which shows the apparent two velocity components near the \NtwoH{} peak (second panel).
This could be explained by radial infall broadening the line near the central region of the dense core, and the apparent two velocity components are the red and blue-shifted emission from the radial infall. 
Nevertheless, we note that the radial infall also could have a rotation component, or the envelope is elongated rather than spherical and thus the anisotropic infall is contributing to the velocity gradient across the source that we measured.
Though the fitted centroid velocity and linewidth shown in Figure~\ref{mom1} are derived assuming single velocity components and could have errors toward the central region of the core where multiple velocity components emerge (e.g.,~locations 1 \& 2 in Figures~\ref{L1451n2hp} \& \ref{L1451spec}), this should not affect our measurement of the velocity gradient (and thus the $J-R$ correlation) because we already excluded regions with broader linewidths through the $\sigma_v$ mask. 
More discussion on spectral fitting with multiple velocity components is included in Appendix~\ref{sec:multiv}.

Our results roughly agree with the observations reported by \cite{2017ApJ...838...60M}, who argued that the broad linewidth is likely caused by non-thermal motions. 
Together with the ``blue bulge'' (blueshifted velocity in the central region of the core) seen in their velocity map, \cite{2017ApJ...838...60M} suggested that these dynamic features are produced by infall motion. The ``blue bulge'' is considered as a signature of infall in optically-thick line emission when the redshifted emission comes only from the outer layers of the envelope and therefore has lower excitation temperature compared to the inner, blueshifted layer behind the central star \citep{1994ApJ...431..767W}, which could result in an asymmetric velocity profile in the PV diagram with blueshifted emission being brighter \citep{1999ARA&A..37..311E}.
% To get self-absorption, you would need warmer infalling material emitting in the interior of the core and then colder material in the outer core absorbing. I am not sure we could completely rule this out, but the lower abundance in the center (likely owing to molecular destruction) would make this less likely as well. 
Though it is not clear from our PV diagrams of L1451-mm, the spectra near the center of the core (locations 1 and 2 in Figure~\ref{L1451spec}) do appear to have a brighter lower-velocity component compared to the higher-velocity one.
%However, though quantitative analysis and model fitting as performed in \cite{2017ApJ...838...60M} is beyond the scope of this paper, we would like to point out that the ``blue bulge'' is not apparent in our observation. 
%In addition, considering the potential lower abundance of \NtwoH{} in warmer gas near the central protostar because of molecular destruction \citep{2004A&A...419L..35B,2010ApJ...708.1002F,2013ApJ...765...18T}, it is less likely that \NtwoH{} is optically-thick in the infall gas.
%{\bf In fact, our fitting results suggest that the total opacity of the hyperfine lines of \NtwoH{} is $\tau_{\rm total} \sim 2-10$ in our data, which means the isolated component (which we used to draw PV diagrams) has opacity $\sim 0.2-1$ and is not optically-thick among our dense core targets. 
We discuss in more detail spectral fitting with multiple velocity components and the corresponding optical depths in Appendix~\ref{sec:multiv}.

\subsection{Per~30}
\label{sec::per30}

\begin{figure*}
\begin{center}
\includegraphics[width=\textwidth]{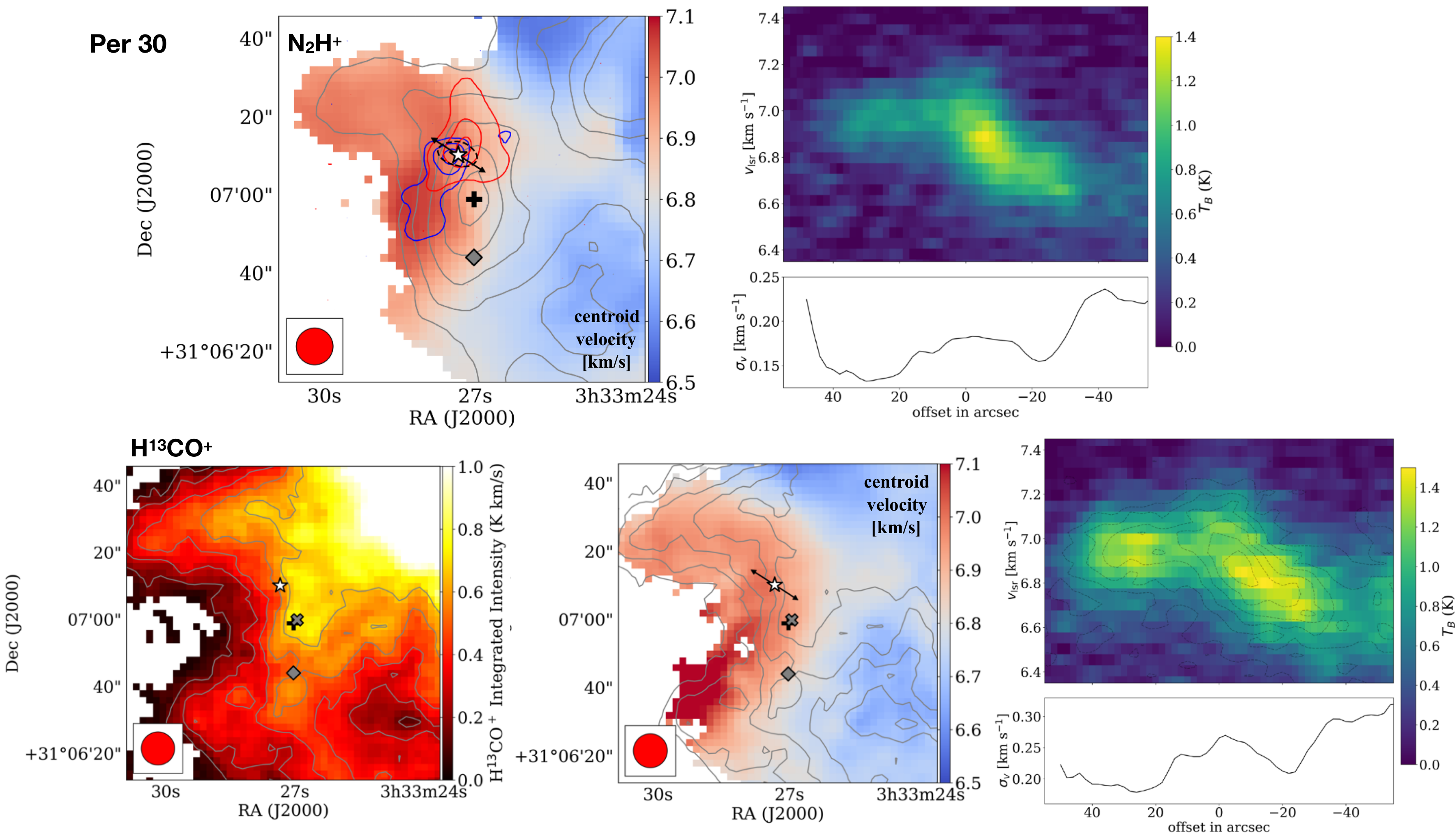}
\vspace{-.15in}
\caption{{\it Top row:} The \NtwoH{} centroid velocity map of Per~30 ({\it left panel}; same as the one in Figure~\ref{mom1}) with \tweCO{} outflows ({\it blue and red contours}; \protect\citealt{2018ApJS..237...22S}), and the PV diagram and P-$\sigma_v$ plot ({\it right panel}) along the velocity gradient ({\it black arrow}) averaged over the region with 1.3~mm emission ({\it dashed contour}; \protect\citealt{2018ApJS..237...22S}).
{\it Bottom row:} Similar as the \NtwoH{} data of Per~30 in Figure~\ref{mom1}, but showing \HthrCO{} data instead. 
The contour levels are integrated \HthrCO{} intensity at $[0.16,0.28,0.4,0.53,0.65]$ K~\kms.
Grey ``x'' signs mark the location of peak \HthrCO{} emission. Note that since the \HthrCO{} data is noisier than \NtwoH, we adopted the direction of velocity gradient measured in \NtwoH{} around the protostar ({\it black arrows in the $v_\mathrm{lsr}$ maps in both rows}) to draw the PV diagram and P-$\sigma_v$ plot for \HthrCO{} ({\it right panel}), which shows two separated clumps of emission on both sides of the protostar. The distribution of \NtwoH{} emission in the PV space centered at the protostar ({\it top right}) is overplotted as grey contours (with levels $T_B = [0.2,0.4,0.6,0.8,1.0,1.2,1.4]$~K) on the PV diagram of \HthrCO{} ({\it bottom right}) for direct comparison.}
\label{Per30H13CO}
\end{center}
\end{figure*}

Per~30 (or Per-emb-30 as denoted in \citealt{2009ApJ...692..973E}) is a Class 0/I protostar with known bipolar outflows traced by \tweCO{} \citep{2018ApJS..237...22S} and a candidate protostellar disk \citep{2018ApJ...866..161S}. 
From the velocity maps and PV diagram shown in Figure~\ref{mom1}, Per~30 seems to be a nice example of velocity structure produced by infall + rotation combined (see e.g.~Figure~1 in \citealt{2012ApJ...748...16T}), with linear velocity gradient at the core scale and slightly increased linewidth toward the center of the core.

However, note that the PV diagram shown in Figure~\ref{mom1} is centered at the \NtwoH{} peak, not the protostar \citep{2009ApJ...692..973E,2016ApJ...818...73T}.  The offset between the \NtwoH{} peak and the protostar is $\gtrsim 10$\arcsec{} ($\approx 0.015$~pc in Perseus), which is almost the size of typical star-forming cores, and is too large to be explained by resolution effect (considering the 7\arcsec{} resolution of {\it Spitzer} at 24~$\mu$m in \citealt{2009ApJ...692..973E} and $< 1$\arcsec{} resolution of VLA in \citealt{2016ApJ...818...73T}), dispersal of the newly formed protostar (typically less than half of the core radius for YSOs; \citealt{2007ApJ...656..293J}), or optical depth. 
Also, though there have been observational evidence of \NtwoH{} being destroyed in warm gas surrounding the central protostar, it likely happens at smaller scales ($\lesssim 0.01$~pc, or 7\arcsec{} at $d\approx 300$~pc; e.g.,~\citealt{2010ApJ...708.1002F,2013ApJ...765...18T}).
All of these make the ``rotation'' feature doubtful. 

In fact, if we re-draw the PV diagram to be centered at the peak continuum emission (determined using $1.3$~mm continuum data from \citealt{2018ApJS..237...22S}) and along the direction of local velocity gradient (calculated within the extended continuum source), we see that the \NtwoH{} core could just be a clump of cold, dense gas sitting on one side of the protostar as projected on the plane of sky (see top panels of Figure~\ref{Per30H13CO}). 
Though bulk proper motion of the entire dense core is unlikely, there are observational examples showing that gas could be flowing toward the protostar along filamentary or asymmetric envelopes (``projected infall''; \citealt{2011ApJ...740...45T,2012ApJ...748...16T}).
Therefore, the \NtwoH{} core could be moving toward the protostar (and thus the linear velocity gradient), or they could be totally unrelated.

The gas structure near the protostar can be more clearly seen in \HthrCO{} data, which is shown in the bottom panels of Figure~\ref{Per30H13CO}.
We see that the \HthrCO{} data does not completely follow the \NtwoH{} emission, nor is it centered around the protostar (bottom left panel in Figure~\ref{Per30H13CO}).
However, the velocity structure is strikingly similar between these two species (see the two $v_\mathrm{lsr}$ maps in Figure~\ref{Per30H13CO}). 
We note that this is similar to the results in \cite{2018A&A...617A..27P}, who also found imperfect spatial correlation between \NtwoH{} and \HthrCO{} within dense cores (their Figure~3), while the velocity profiles traced by these two species are highly similar (their Figure~5). Similar results were also reported in filamentary structures using \NtwoH{} and C$^{18}$O by \cite{2011A&A...533A..34H}.

The fact that core materials with different densities have similar kinematic features
could suggest that the velocity field in this region corresponds to larger-scale motions and is not induced by the local rotation/infall around the protostellar system or the \NtwoH{} core.
In addition, from the PV diagram of \HthrCO{} centered at the protostar (bottom right panel in Figure~\ref{Per30H13CO}), there are clearly two spatially-separated clumps of emission on the two sides of the protostar. 
Whether or not these \HthrCO{} clumps are real structures, and whether they contribute to the formation of the protostar, remains uncertain based on the current data. 

Another feature we would like to point out is the \NtwoH{} linewidth near the \NtwoH{} core (see the middle and right panels of the Per~30 row in Figure~\ref{mom1}). The P-$\sigma_v$ profile shows that there are ``dips'' of $\sigma_v$ on the edges of the core, forming a ring-like low-$\sigma_v$ region outside the \NtwoH{} core, inside which the linewidth increases again toward the center of the core. The increased linewidth near the center of the \NtwoH{} core is similar to what we observed in L1451-mm (see Section~\ref{sec::l1451mm} above), which could be explained by infalling gas. However, this becomes less clear when considering that the \NtwoH{} emission is far offset from the protostar, which should have the dominant gravitational field in this region.\footnote{This would of course depend on the mass of the protostar compared to that of the \NtwoH{} core. However, assuming the protostar and the core co-exist in the same local region of the MC, the fact that the protostar is at a later evolutionary stage than the core indicates that the protostar is likely more massive. In fact, a rough estimate using {\it Herschel} column density map gives the mass of the dense core $\sim 0.4$~M$_\odot$, while the mass of the protostellar disk around the protostar has been reported to be $\sim 0.1-0.2$~M$_\odot$ in \cite{2016ApJ...817L..14S}. Considering that a protostar is typically much more massive than its disk, this agrees with our initial guess that the gravity of the protostar dominates in this region.} At this point, it is not clear that whether the linewidth profile is providing important dynamical information or is affected by optical depth. This will be examined in our follow-up studies using detailed radiative transfer modeling to study the spectral line properties within collapsing/rotating cores.

\subsection{Starless cores B1-SW and B1-NE}
\label{sec::starless}

Two of our targets, B1-SW and B1-NE, are dense, starless cores located in the Barnard~1 region of the Perseus MC previously identified both in $1.1$~mm continuum \citep[see Table~\ref{tab::obs};][]{2006ApJ...638..293E} and \NtwoH{} emission \citep[CLASSy;][]{2014ApJ...794..165S}. 
They are classified as starless cores mainly because of the lack of point-source detection in the mid-infrared \citep{2006ApJ...638..293E,2009ApJ...692..973E}.

Looking at the centroid velocity maps in Figure~\ref{mom1}, we clearly see that both starless cores have smooth velocity structures, with obvious velocity gradients over the observed regions. 
However, we first noticed that in B1-SW, the velocity gradient is not monotonic: the gas becomes more redshifted toward the bright center of \NtwoH{} emission from both east and west halves of the core. This is also reflected by the fact that the averaged velocity gradients with and without the narrow-line limit ($\sigma_v < \langle\sigma_v\rangle$) are almost perpendicular to each other (see the black and grey arrows in Figure~\ref{mom1}), because the strongest velocity gradients within the core (east to west on the east side, and west to east on the west side) are easily canceled out and contribute little to the averaged value.
This divergent gradient can be seen more clearly in the PV diagram, and B1-SW appears to have a wedge-like structure in the PV space with the brightest \NtwoH{} emission located at the top (see the right panel of the bottom row in Figure~\ref{mom1}).

Intriguingly, we found a similar feature in B1-NE. Though the velocity structure within the \NtwoH-defined core of B1-NE seems to be a smooth gradient along a single direction (northwest to southeast), toward the northwest slightly outside the core the velocity becomes redshifted, same as the southeast half of the core (see the left panel in the third row of Figure~\ref{mom1}). This again produces a two-side velocity gradient which can be clearly seen in the PV diagram (see the right panel of the third row in Figure~\ref{mom1}). Similar as but slightly different from the case of B1-SW, the velocity structure around the B1-NE core looks like a downward arc with an offset \NtwoH{} peak on the left side of the arc. 

These features suggest that both of the starless cores in our observations are located near local extremes in $v_\mathrm{lsr}$, which is in good agreement with the results found in some of the pressure-confined cores (``droplets'') using NH$_3$ in the L1688 region of the Ophiuchus molecular cloud \citep{2018arXiv180910223C}.
This could imply that starless cores form via gas flows that collide obliquely at the location where flow-compressed materials eventually become dense cores, a process seen in various numerical simulations \citep[see e.g.,][]{2000prpl.conf....3V,2001ApJ...553..227P,2007prpl.conf...63B}.
The origin of dense cores as a result of accretion shocks has also been discussed in \cite{2010ApJ...712L.116P} because of the observed sharp transition to coherence.
A rough, order-of-magnitude estimate with the assumption that the momentum lost during the shock is isotropic would indicate a shock compression ratio of $\rho_{\rm post-shock}/\rho_{\rm pre-shock} \sim {\cal M}^2 \sim (\Delta v/c_s)^2$. Since $\Delta v_{\rm los} \approx 0.4$~\kms{} for both B1-NE and B1-SW (see the PV diagrams in Figure~\ref{mom1}), we have $\rho_{\rm post-shock}/\rho_{\rm pre-shock} \sim 4$ assuming $c_s = 0.2$~\kms. The {\it Herschel} column density data shows a factor of $\sim 4-5$ enhancement in $N_{\rm H}$ across the cores, which roughly agree with the shock compression ratio if we further assume the cores are formed within a locally-flat region, i.e.,~$\Sigma_{\rm post-shock}/\Sigma_{\rm pre-shock} \sim \rho_{\rm post-shock}/\rho_{\rm pre-shock}$.

Indeed, we cannot fully rule out the possibility of gravity-induced, magnetic field-regulated (i.e.,~anisotropic) density enhancement within these cores. 
More detailed analytical analysis, numerical modeling, and synthetic observations are underway to further investigate these two scenarios and will be discussed in a follow-up paper.
Future survey-style observations will also be helpful to statistically determine if the divergent velocity gradient and the arc-shape PV structure revealed in this dataset are common among starless cores.

\subsection{IRAS7}
\label{sec::iras7}

\begin{figure}
\begin{center}
\includegraphics[width=0.9\columnwidth]{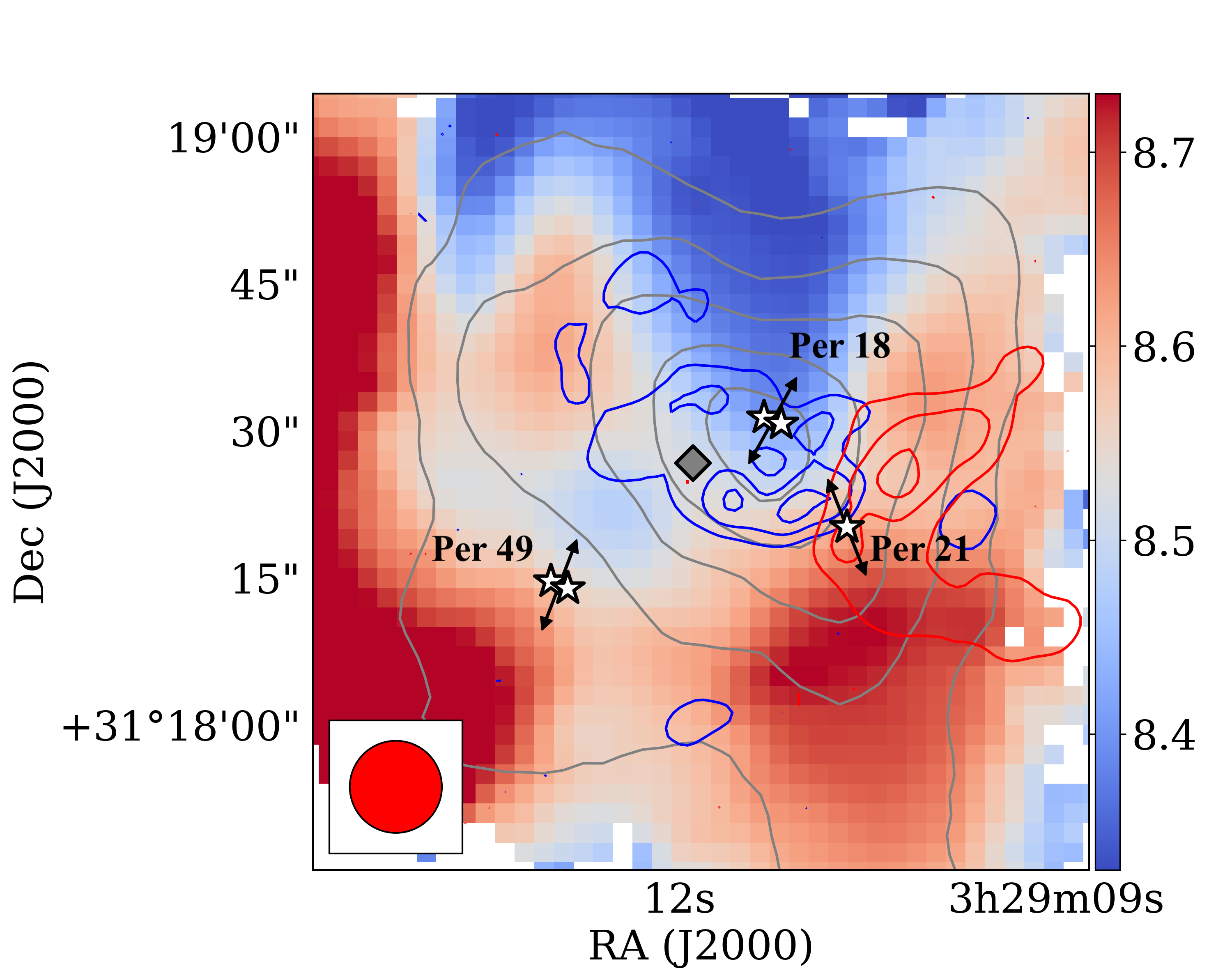}
\vspace{-.1in}
\caption{Illustration of the complex structure within IRAS7, showing the centroid velocity (in \kms) traced by \NtwoH{} ({\it colormap}) with fitted local velocity gradients ({\it black arrows}) same as in Figure~\ref{mom1IRAS7}. Protostars are marked as white stars, and the outflow (mostly dominated by Per~21) traced by \tweCO{} \protect\citep{2018ApJS..237...22S} is included ({\it blue and red contours}). Outflow from Per~18 has also been detected and is nearly north-south \citep{2018ApJ...867...43T}. The separations of the binary systems in Per~18 and Per~49 are for demonstration only and are not drawn to scale. }
\label{IRAS7outflow}
\end{center}
\end{figure}

IRAS7 is a dense core in the NGC~1333 region of Perseus \citep{1987MNRAS.226..461J, 1998A&A...334..269L} which appears to be a triple system in {\it Spitzer} images \citep{2009ApJ...692..973E}.
The three protostellar components of the triple system, Per~18, Per~21, and Per~49, were later included in the VLA Nascent Disk and Multiplicity (VANDAM) survey to study the inner-most regions surrounding the protostellar sources \citep{2016ApJ...818...73T} and the Mass Assembly of Stellar Systems and their Evolution with the SMA (MASSES) survey to study gas dynamics of outflows and protostellar envelopes \citep{2016ApJ...820L...2L,2017A&A...602A.120F,2018ApJS..237...22S,2019ApJ...873...54A}. 

A complete view of the velocity field traced by our \NtwoH{} data is shown in Figure~\ref{IRAS7outflow}, with \tweCO{} outflows (most likely from Per~21) from the MASSES survey overplotted. 
Note that \tweCO{} outflows from Per~18 were also detected at smaller scales ($\lesssim 3$\arcsec), which is nearly north-south and edge-on \citep{2018ApJ...867...43T}. 
Among all 5 targets observed in this study, IRAS7 has the most complex velocity field (see Figures~\ref{mom1} and \ref{IRAS7outflow}), which can be directly linked to the clustered environment of IRAS7. 
Despite the disordered velocity structure at the core scale, we actually found that the three protostellar systems in IRAS7 all have simple, one-directional velocity gradients locally within the $\sim 20$\arcsec{} scale (see Figure~\ref{mom1IRAS7}) that, when interpreted as rotation, fit in the specific angular momentum$-$radius correlation very well (Figure~\ref{JRcorr}). In addition, the local velocity fields around the three protostellar systems seem to be inter-connected spatially (Figure~\ref{IRAS7outflow}). This is especially obvious for the pairs Per~18$-$Per~21 and Per~18$-$Per~49 by looking at the gas structures in the PV diagrams (Figure~\ref{mom1IRAS7}), where we see the brightest \NtwoH{} emission within IRAS7 (originated around Per~18) is connected to local dense gas surrounding both Per~21 and Per~49 in the PV space.

Interestingly, the two protostellar systems in IRAS7, Per~18 and Per~49, are multiple systems themselves \citep{2016ApJ...818...73T}, as illustrated in Figure~\ref{IRAS7outflow} (not to scale). Surprisingly, these two multiple systems both have local velocity gradients perpendicular to the projected separation between the binaries.
Assuming the velocity gradient is representing the rotation of local gas, 
this is completely opposite to what described by the classical star formation theory,\footnote{The measurable quantity here is the orientation of the line connecting the two stellar components in the sky plane. The angle between this line and the angular momentum axis of the orbit (perpendicular to the orbital plane in 3D) projected in the sky plane can have any value, depending where the secondary is in its orbit, which has the shape of an ellipse in the plane of the sky for an intrinsically circular orbit. In particular, when the companion is at a location along the minor axis of the (projected) ellipse, the orientation of the two stars in the sky plane would be the same as the orientation of the orbital angular momentum in the sky plane, which would be perpendicular to the orientation of velocity gradient. So it is plausible although unlikely that the orbital angular momentum and the velocity gradient could be consistent. The chance for one system for the secondary to locate near the minor axis of the (projected) orbit is already small. That for both to do the same is smaller.} which envisions the flattening of dense core as the consequence of angular momentum conservation, and any sequential evolution of structures (protostellar disk, multiple system, etc.) should all take place in the plane perpendicular to the rotational axis of the core \citep[for a review, see e.g.,][]{2002ARA&A..40..349T}.
This could suggest that the rotation at core/envelope scales does not transfer efficiently all the way to the disk-forming scale, or the velocity gradient observed here is not generated by rotation. 
In fact, a recent study by \cite{2018ApJ...867...43T} has shown that gas rotation in Per~18 at small scales ($\sim 1$\arcsec; see their Figure~10) is orthogonal to the local velocity gradient that we observed.
However, it is worth mentioning that the binary-forming plane could also be altered by other mechanisms (e.g.,~turbulent accretion; interaction between forming multiple companions and local turbulence within the core) at later stages during disk formation \citep{2015Natur.518..213P,2016ApJ...827L..11O}.
Future studies on scales between 1\arcsec$-20$\arcsec{}
%with higher angular resolution 
will help investigate the progression of gas kinematics from outer envelopes to disk-forming regions.
%distinguish these two scenarios.

\section{Summary}
\label{sec::sum}

Velocity information within dense cores is one of the critical properties needed to understand the star forming process, because it provides the initial conditions for all protostellar evolution. However, there is yet not much investigation on core-scale kinematics simply because of a lack of the required combination of spatial dynamic range, spectral resolution, and sensitivity for appropriate dense core molecular line tracers. Here we report the high-spectral resolution \NtwoH{} J=1-0 line data toward 5 dense cores (and \HthrCO{} data of Per~30) in the Perseus MC with 9\arcsec{} resolution and $\gtrsim 1$\arcmin{} spatial coverage using the {\it Argus} focal plane array on GBT. 

We summarize our main conclusions below:
\begin{enumerate}
\item Spectral resolution is crucial for revealing the true dynamic features within dense cores. Though the fitted centroid velocity field may seem smooth and show a clear gradient that can be interpreted as rotation (Figure~\ref{JRcorr}), the detailed gas structure in the position-velocity space, which is only achievable with a high-enough spectral resolution, can provide better constraints on the true nature of gas kinematics in these star progenitors (Figures~\ref{mom1} and \ref{mom1IRAS7}). 

\item Our data on L1451-mm roughly agrees with the infall scenario discussed in \cite{2017ApJ...838...60M}, with 
%brighter emission on the blueshifted gas near the center of the core and 
spectra showing multiple velocity components toward the center of the core (Figure~\ref{L1451spec}) and
the significantly larger linewidth within the dense core (top row of Figure~\ref{mom1}; also see Figure~\ref{L1451n2hp}). Moreover, the extended \NtwoH{} emission on the northwest side of the core and the smooth, coherent velocity field over the $\gtrsim 1$\arcmin{} ($\sim 0.08$~pc) gas structure suggest that the core-scale kinematics are connected to the cloud-scale gas motion.

\item Despite the well-ordered velocity field in the region, we found that the \NtwoH{} emitting dense core near the Class 0/I protostar Per~30 \citep{2016ApJ...818...73T} may not be dynamically tied to the protostar, considering the separation between the \NtwoH{} peak and the protostar is wide ($\gtrsim 10$\arcsec) and the gas structure in the PV space seems to center on the \NtwoH{} peak instead of the protostar (Figure~\ref{mom1}). Additional \HthrCO{} data around Per~30 also show two separated clumps in the PV space on both sides of the protostar (Figure~\ref{Per30H13CO}), but whether or not this is a physical or chemical effect remains uncertain. 

\item Both of the two starless cores observed in this study, B1-NE and B1-SW, show arc-like structures in the PV space (Figure~\ref{mom1}), suggesting that they may have formed near the convergent point of two oblique gas flows, a feature of the so-called turbulent model of star formation \citep[for reviews, see][]{2000prpl.conf....3V,2007prpl.conf...63B}.
This is also in good agreement with the results found in pressure-confined clumps in the L1688 region in Ophiuchus, that these clumps tend to be located at local extrema of centroid velocity \citep{2018arXiv180910223C}.

\item As a triple system-forming dense core, IRAS7 shows the most complex velocity structure at the core scale (Figure~\ref{IRAS7outflow}), though locally ($\lesssim 20$\arcsec) there are simple, one-directional velocity gradients around each of the protostellar systems forming in IRAS7. Interestingly, the local velocity gradients near the two multiple systems in IRAS7, Per~18 and Per~49, are both nearly perpendicular to the projected separation between the binaries \citep{2016ApJ...818...73T}. 
Moreover, the gas rotation in the pseudo-disk/inner-most envelope of Per~18 ($\lesssim 1$\arcsec) also appears to be orthogonal to the 20\arcsec-scale velocity gradient \citep{2018ApJ...867...43T}.
This could suggest that the local velocity gradient is not induced by rotation, or that the rotation at the envelope scale is not efficiently transported to the disk-forming scale.

\item The kinematic features revealed in this study are intriguing, but many fundamental questions remain uncertain with only 5 targets observed. Future survey-style observational projects will help provide answers statistically on the origin of angular momenta within star-forming cores and the connection between core-scale and cloud-scale gas dynamics.

\end{enumerate}

\section*{Acknowledgements}

We thank the referee for a very helpful report.
The authors would like to thank the {\it Argus} instrument team from the Stanford University, Caltech, JPL, University of Maryland, University of Miami, and the Green Bank Observatory for their efforts on the instrument and software that have made this work possible. 
The {\it Argus} instrument construction was funded by NSF ATI-1207825.
Green Bank Observatory is a facility of the National Science Foundation and is operated by Associated Universities, Inc.
CYC and ZYL acknowledge support from NSF AST-1815784. ZYL is supported in part by NASA 80NSSC18K1095 and NNX14AB38G and NSF AST-1716259.
AIH and JL acknowledge support from NSF AST-1615647.
This research made use of Astropy,\footnote{http://www.astropy.org} a community-developed core Python package for Astronomy \citep{2013A&A...558A..33A,2018AJ....156..123A}.

%%%%%%%%%%%%%%%%%%%%%%%%%%%%%%%%%%%%%%%%%%%%%%%%%%



%\bibliographystyle{mnras}


%%%%%%%%%%%%%%%%%%%%%%%%%%%%%%%%%%%%%%%%%%%%%%%%%%

%%%%%%%%%%%%%%%%% APPENDICES %%%%%%%%%%%%%%%%%%%%%

\appendix

\section{Spectral Fitting Test with Multiple Velocity Components}
\label{sec:multiv}

\begin{figure}
\begin{center}
\includegraphics[width=\columnwidth]{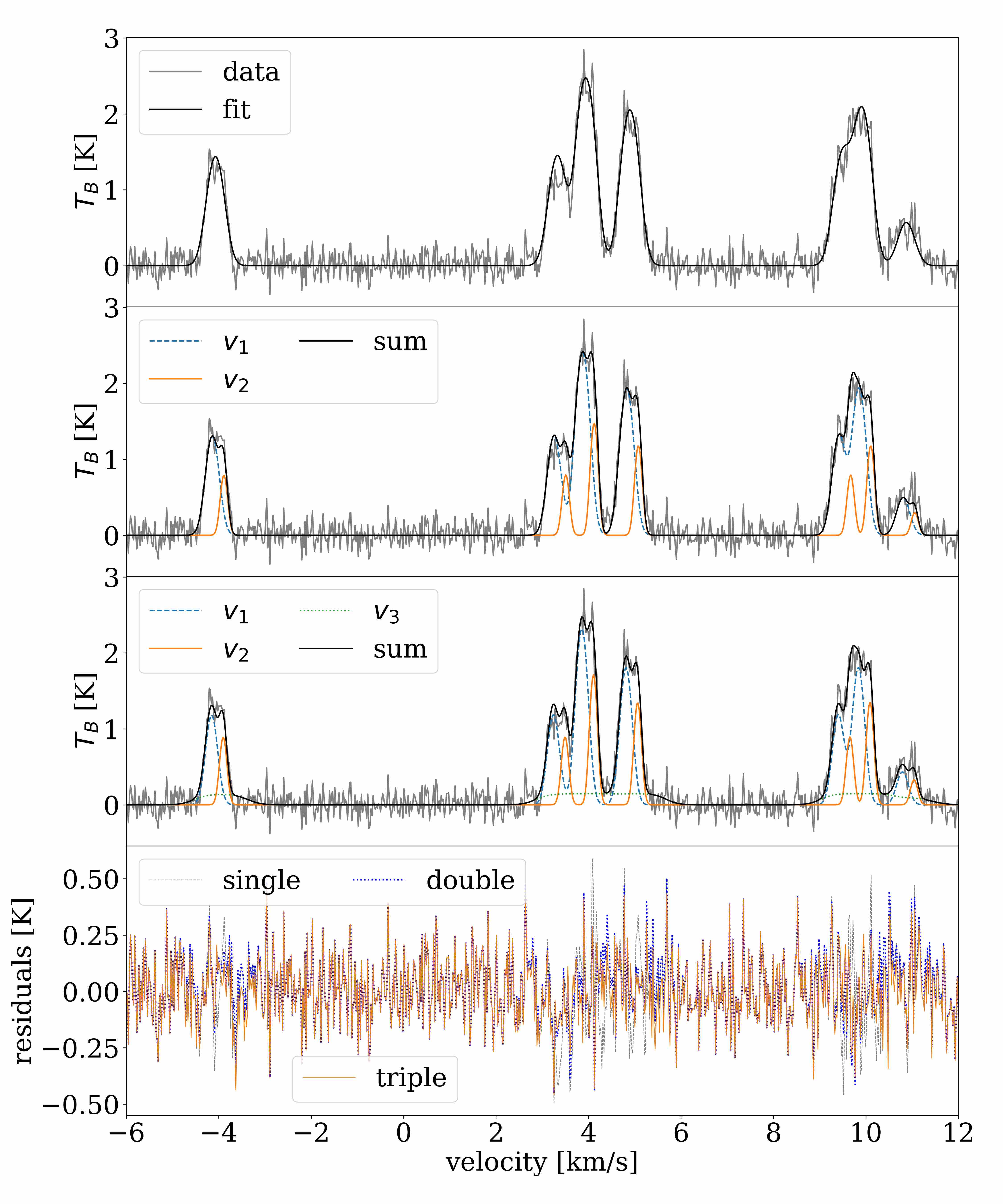}
\vspace{-.15in}
\caption{{\it Top 3 panels:} comparison between spectral fitting results ({\it black}) of our GBT-Argus data ({\it gray}) at location 2 of L1451-mm (the peak of \NtwoH{} integrated intensity; see Figures~\ref{L1451n2hp} and \ref{L1451spec}) assuming single ({\it top row}), double ({\it second row}), and triple ({\it third row}) velocity components. {\it Bottom panel:} the residual from the fits. Though the difference is small, the data is indeed better described by a multi-component emission line. }
\label{L1451multi}
\end{center}
\end{figure}

\begin{table*}
	\centering
  \begin{threeparttable}
	\caption{Spectral fitting results at location 2 of L1451-mm (the peak of \NtwoH{} integrated intensity; see Figure~\ref{L1451n2hp}) with single, double, and triple velocity components.}
\label{tab:multiv}
\begin{tabular}{ l | c c c c c}
\hline
peaks & $T_\mathrm{ex}$~[K] & $\tau_{\rm total}$ & $v_{\rm centroid}$~[\kms] & $\sigma_v$~[\kms] & residual RMS\\
\hline
  single & $5.75\pm 0.19$ & $4.87\pm 0.51$ & $3.967\pm 0.003$ & $0.179 \pm 0.003$ & 0.1312 \\
  \hline
  \multirow{2}{*}{double} & $6.17\pm 0.34$  & $3.65\pm 0.53$ & $3.898\pm 0.008$ & $0.140 \pm 0.006$ & \multirow{2}{*}{0.1196}\\
   &  $4.98\pm 0.61$ & $3.4 \pm 1.2$ & $4.150\pm 0.005$ & $0.075\pm 0.005$  \\
  \hline
  \multirow{3}{*}{triple} & $6.8\pm 0.7$  & $2.57\pm 0.59$ & $3.884\pm 0.007$ & $0.125 \pm 0.006$ & \multirow{3}{*}{0.1166} \\
   &  $5.56\pm 0.73$ & $2.94 \pm 0.99$ & $4.134\pm 0.006$ & $0.077\pm 0.004$  \\
   &  $2.87\pm 0.03$ & $23 \pm 15$ & $4.090\pm 0.064$ & $0.347\pm 0.065$  \\
   \hline
	\end{tabular}
  \end{threeparttable}
\end{table*}

To better understand the gas dynamics towards regions that potentially show features of infall motions (i.e.,~multiple velocity components along the line of sight), we perform spectral fitting at location 2 of L1451-mm (the peak of \NtwoH{} integrated intensity; see Figures~\ref{L1451n2hp} and \ref{L1451spec}) with multiple (2 and 3) velocity components (an optional function of \texttt{PySpecKit}), and compare it with the original fitting result assuming a single velocity component (Figure~\ref{L1451multi}). The fitting results are summarized in Table~\ref{tab:multiv}. 
We found that though the difference is small (see the residuals in the bottom panel of Figure~\ref{L1451multi}), both of the multiple-component fits successfully return two velocity components with offset $\sim 0.15$~\kms, consistent with what we see in the PV diagrams (see Figure~\ref{L1451n2hp}). 
The linewidths of these two velocity components are both smaller than the linewidth in the original fit, which suggests that the broader linewidths in this region when fitted with single velocity component (see Figure~\ref{mom1}) could be caused by overlapping multiple velocity components. 
We also noted that the fitted excitation temperature $T_{\rm ex}$ and total optical depth $\tau_{\rm total}$ of these two velocity components are very close to the single-component fit.

More importantly, the fitting results agree with the ``blue bulge'' feature suggested by \cite{2017ApJ...838...60M}, that the blue-shifted component is brighter (higher excitation temperature). This motivated us to include the third velocity component in the fit (third panel in Figure~\ref{L1451multi}) to describe the emission from central region, which is expected to be optically-thick in the ``blue bulge'' picture. 
We found that including this optically-thick, very broad emission line with centroid velocity in between the redshifted and blue-shifted components does return a slightly better fit to the data by reducing the residual value down by $\sim 3\%$ of the value corresponding to the 2-component fit. 
Though preliminary, this could provide important information when investigating the potential infall profile toward the center of the core.
More detailed spectral analysis is underway and will be discussed in a separate paper.

%\section{Dense Gas At An Arced Shock Front}
%\label{sec:bow}

%\begin{figure}
%\begin{center}
%\includegraphics[width=0.6\columnwidth]{bowshock.png}
%\vspace{-.15in}
%\caption{\bf test }
%\label{fig:bow}
%\end{center}
%\end{figure}

%From Larson's law,
%\begin{equation}
%v_0 \approx \alpha R^{1/2}.
%\end{equation}
%Therefore
%\begin{equation}
%\begin{split}
%R &= \left(\frac{v_0}{\alpha}\right)^2 = \left(\frac{\Delta v}{1-\cos\theta} \frac{1}{\alpha}\right)^2\\
%&= \frac{\Delta s}{\cos\theta},
%\end{split}
%\end{equation}
%or
%\begin{equation}
%\frac{\cos\theta}{\left(1-\cos\theta\right)^2} = \frac{\Delta s \alpha^2}{\Delta v^2}.
%\end{equation}

%%%%%%%%%%%%%%%%%%%%%%%%%%%%%%%%%%%%%%%%%%%%%%%%%%

% Don't change these lines
\bsp	% typesetting comment
\label{lastpage}
\end{document}